\documentclass[twocolumn]{aastex631}
\usepackage{amsmath}
\usepackage{float}

\usepackage{appendix}
\usepackage{multirow}
\usepackage{hhline}
\usepackage{longtable}

\received{November 16, 2022}
\revised{March 6, 2023}
\accepted{March 6, 2023}

\shorttitle{The recalibrated UVES-POP spectral library}
\shortauthors{Borisov et al.}

\begin{document}

\title{New Generation Stellar Spectral Libraries in the Optical and Near-Infrared I:\\
The Recalibrated UVES-POP Library for Stellar Population Synthesis\footnote{Recalibrated spectra are available at \url{https://sl.voxastro.org/}}}

\correspondingauthor{S. Borisov; I. Chilingarian}

\author[0000-0002-2516-9000]{Sviatoslav B. Borisov}
\affil{Department of Astronomy, University of Geneva, Chemin Pegasi 51, 1290 Versoix, Switzerland}
\affil{Sternberg Astronomical Institute, Moscow M.V. Lomonosov State University, 13 Universitetsky pr., Moscow 119234, Russia}
\email{sviatoslav.borisov@unige.ch; igor.chilingarian@cfa.harvard.edu}

\author[0000-0002-7924-3253]{Igor V. Chilingarian}
\affil{Center for Astrophysics -- Harvard and Smithsonian, 60 Garden Street MS09, Cambridge, MA 02138, USA}
\affil{Sternberg Astronomical Institute, Moscow M.V. Lomonosov State University, 13 Universitetsky pr., Moscow 119234, Russia}

\author[0000-0001-8427-0240]{Evgenii V. Rubtsov}
\affil{Sternberg Astronomical Institute, Moscow M.V. Lomonosov State University, 13 Universitetsky pr., Moscow 119234, Russia}
\affil{Faculty of Physics, Moscow M.V. Lomonosov State University, 1 Leninskie Gory, Moscow 119991, Russia}

\author[0000-0002-7864-3327]{C\'edric Ledoux}
\affil{European Southern Observatory, Alonso de C´ordova 3107, Vitacura, Casilla 19001, Santiago de Chile, Chile}

\author[0000-0002-6090-8446]{Claudio Melo}
\affil{Portuguese Space Agency, Estrada das Laranjeiras, n .$^{\circ}$205, RC, 1649-018, Lisboa, Portugal}

\author[0000-0003-3255-7340]{Kirill A. Grishin}
\affil{Universit\'e de Paris-Cit\'e, CNRS/IN2P3, Astroparticule et Cosmologie, F-75013 Paris, France}
\affil{Sternberg Astronomical Institute, Moscow M.V. Lomonosov State University, 13 Universitetsky pr., Moscow 119234, Russia}

\author[0000-0002-6425-6879]{Ivan Yu. Katkov}
\affil{New York University Abu Dhabi, Saadiyat Island, PO Box 129188, Abu Dhabi, UAE}
\affil{Sternberg Astronomical Institute, Moscow M.V. Lomonosov State University, 13 Universitetsky pr., Moscow 119234, Russia}

\author[0000-0002-2550-2520]{Vladimir S. Goradzhanov}
\affil{Sternberg Astronomical Institute, Moscow M.V. Lomonosov State University, 13 Universitetsky pr., Moscow 119234, Russia}
\affil{Faculty of Physics, Moscow M.V. Lomonosov State University, 1 Leninskie Gory, Moscow 119991, Russia}

\author[0000-0002-8220-0756]{Anton V. Afanasiev}
\affil{LESIA, Observatoire de Paris, Universit\'e PSL, CNRS, Sorbonne Universit\'e, Universit\'{e} Paris Cit\'{e}, 5 place Jules Janssen, 92195, Meudon, France}
\affil{Sternberg Astronomical Institute, Moscow M.V. Lomonosov State University, 13 Universitetsky pr., Moscow 119234, Russia}

\author[0000-0002-1091-5146]{Anastasia V. Kasparova}
\affil{Sternberg Astronomical Institute, Moscow M.V. Lomonosov State University, 13 Universitetsky pr., Moscow 119234, Russia}

\author[0000-0002-4342-9312]{Anna S. Saburova}
\affil{Sternberg Astronomical Institute, Moscow M.V. Lomonosov State University, 13 Universitetsky pr., Moscow 119234, Russia}

\begin{abstract}
We present re-processed flux calibrated spectra of 406 stars from the UVES-POP stellar library in the wavelength range 320--1025 nm, which can be used for stellar population synthesis. The spectra are provided in the two versions having spectral resolving power $R=$20,000 and $R=$80,000. Raw spectra from the ESO data archive were re-reduced using the latest version of the UVES data reduction pipeline with some additional algorithms that we developed. The most significant improvements in comparison with the original UVES-POP release are: (i) an updated Echelle order merging, which eliminates ``ripples'' present in the published spectra, (ii) a full telluric correction, (iii) merging of non-overlapping UVES spectral setups taking into account the global continuum shape, (iv) a spectrophotometric correction and absolute flux calibration, and (v) estimates of the interstellar extinction. For 364 stars from our sample, we computed atmospheric parameters $T_{\mathrm{eff}}$, surface gravity log $g$, metallicity [Fe/H], and $\alpha$-element enhancement [$\alpha$/Fe] by using a full spectrum fitting technique based on a grid of synthetic stellar atmospheres and a novel minimization algorithm. We also provide projected rotational velocity $v \sin i$ and radial velocity $v_{rad}$ estimates. The overall absolute flux uncertainty in the re-processed dataset is better than 2\%\ with sub-\%\ accuracy for about half of the stars. A comparison of the recalibrated UVES-POP spectra with other spectral libraries shows a very good agreement in flux; at the same time, \textit{Gaia} DR3 BP/RP spectra are often discrepant with our data, which we attribute to spectrophotometric calibration issues in \textit{Gaia} DR3.
\end{abstract}


\section{Introduction} \label{sec:intro}

High quality models of stellar populations play a crucial role in modern astrophysics -- they are used for interpreting galaxy and star cluster data to determine the age, metallicity, chemical abundances, and other properties \citep[e.g., ][]{vazd_1,delgado,bruz_char,gui_roc}. For the vast majority of stellar systems, only integrated light characteristics are available such as colors and/or spectra integrated along the line of sight, that is a sum of contributions from individual stars, which are too faint and/or too crowded to be observed single-handedly. Spectra of synthetic stellar populations are used to interpret such datasets and infer star formation and chemical enrichment histories of galaxies and their subsystems. An essential ingredient of stellar population synthesis is a library of stellar spectra which is a set of spectra of stars in some particular wavelength range and having the same spectral resolution. Stars comprising a spectral library should have the widest possible coverage in physical parameters of stellar atmospheres, such as the effective temperature $T_{\mathrm{eff}}$, surface gravity log~$g$, iron abundance [Fe/H], $\alpha$-elements enhancement [$\alpha$/Fe] (and sometimes abundances of individual chemical elements, micro- and macro-turbulent velocities etc.)

In this paper, we describe the recalibration process of the UVES-POP stellar spectral library (UVES Paranal Observatory Project; \citealp{uves_pop_orig}) which includes spectra of 406 stars observed with the Ultraviolet and Visual Echelle Spectrograph (UVES; \citealp{dekker00}) operated at ESO~VLT~UT2. The motivation for our project was that most empirical fully calibrated stellar spectra have either high resolution but rather short wavelength range, or wide spectral coverage but low-to-intermediate resolution. The resolving power of the UVES-POP spectra is R$\sim$80,000 whilst covering the wavelength range from 304~nm to 1040~nm almost contiguously. The main goals of the UVES-POP library recalibration project are to improve the merging of Echelle orders, to perform telluric correction, spectrophotometric calibration and correction for interstellar extinction, so as to make spectra usable for stellar population synthesis. In addition, we present atmospheric parameters of UVES-POP stars computed with a novel full spectrum fitting technique.

\begin{deluxetable*}{lcccc}
\tabletypesize{\footnotesize}
\tablewidth{3pt}
\tablecaption{Characteristics of several widely-used stellar libraries. The information about UVES-POP refers to the original release.\label{stellar_libs}}
\tablehead{
\colhead{Library} & \colhead{Resolving	power R} & \colhead{Wavelength range, nm} & \colhead{Number of stars} & \colhead{Reference}
}
\startdata 
UVES-POP	    & 80,000      		& $304-1040$      		& 359	&   \citet{uves_pop_orig}	\\
SpecMatch-Emp   & 60,000$^*$        & $499-641$             & 404   &   \citet{FGKM_library}    \\
ELODIE		    & 10,000    		& $390-680$      		& 1388  &   \citet{elodie_orig},    \\
                & 42,000$^*$        & $390-680$             & 1388  &   \citet{elodie_new_2}    \\
X-Shooter	    & 10,000      		& $300-1020$      		& 237	&	\citet{x_shooter},		\\
                & 10,000            & $300-2450$            & 666   &   \citet{x-shooter-dr2} (DR2),  \\
                & 10,000            & $350-2480$            & 683   &   \citet{x-shooter-dr3} (DR3)   \\
Indo-US		    & 5,000     	    & $346-946$     		& 1273	&   \citet{indo-us}		    \\
MILES 		    & 2,000      		& $352-750$      		& 985	&   \citet{miles}			\\
STELIB 		    & 2,000      		& $320-930$      		& 249	&   \citet{stelib}			\\
IRTF            & 2,000             & $800-5000$            & 210   &   \citet{irtf_dr1}        \\
                & 2,000             & $700-2500$            & 284   &   \citet{irtf_dr2}        \\
MaStar          & 1,800             & $362-1035$            & 3321  &   \citet{manga_library}   \\
LW2000		    & 1,100     	    & $500-2500$     		& 100	&   \citet{lw2000}		    \\
NGSL            & 1,000             & $168-1020$            & 378   &   \citet{ngsl}            \\
\enddata
\tablecomments{$^*$ continuum normalized spectra}
\end{deluxetable*}

\section{The original UVES-POP stellar library: the sample}
\label{sec:targ}

\begin{figure}
\centering
\includegraphics[width=1\linewidth]{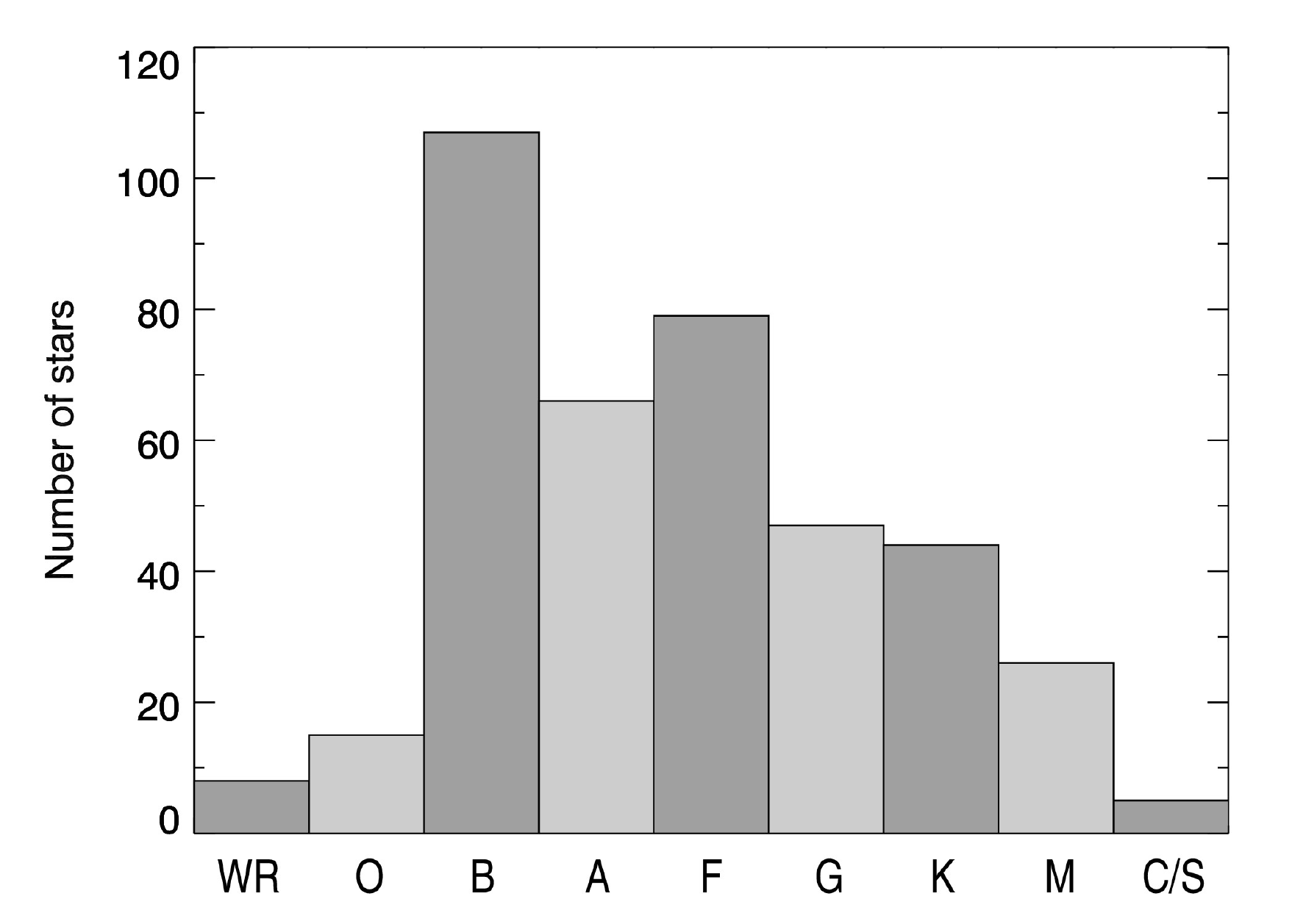}
\caption{Distribution of stellar spectral types in the UVES-POP library. Spectral type identification was done mostly using the SIMBAD database (\citealt{simbad}) or, in rare cases, based on stellar atmospheric parameters (see section~\ref{sec:atmpar}).}
\label{sp_types}
\end{figure}

The original UVES-POP stellar spectral library contains spectra of around 400 stars with a resolving power of R=80,000. The stars were observed with UVES using a 0.5$''$-wide slit in two instrument modes, combining Dichroic~\#1 and Dichroic~\#2, in order to cover the wavelength range from 300 to 1040~nm contiguously with only a few narrow gaps. Observations were carried out between February 2001 and March 2003 under ESO program 266.D-5655(A). We selected all stars from the UVES ESO archive for that program and downloaded the original unprocessed data. We also downloaded calibration frames collected as close as possible in time to the science frames. In our sample, there are several targets that were not included in the final sample of the original UVES-POP library because they did not pass quality control. In total, the new recalibrated UVES-POP library contains spectra of 406 stars, 262 of which have spectra covering the full optical wavelength range. Fig.~\ref{sp_types} presents the distribution of stellar types of the targets, which we used in our project.
    
The UVES-POP library includes two groups of stars: field stars and stars in open clusters. The former ones were selected to cover the largest possible variety of stellar types across the Hertzsprung--Russell Diagram. Stars from the second group belong to the two young open clusters IC~2391 (Age $\approx40$ Myr, \citealt{ic2391_age}) and NGC~6475 (Age~$\approx200$~Myr, \citealt{ngc6475_age}), so that the stars in each cluster should have similar chemical abundances. 

\section{Data reduction and calibration}
\label{sec:calib}
\subsection{Preliminary reduction with the UVES pipeline}
\label{subsec:prelim}

We used the UVES pipeline (\citealt{uves_pipeline}) through EsoRex (ESO Recipe Execution Tool; \citealt{esorex}) to process the raw 2D UVES data which include science and calibration frames (bias, format check file, order definition flat, Echelle flat, dark, and a spectrum of the ThAr wavelength calibration lamp). All of them were taken on the same night as the corresponding science frames, except for the dark frames which were taken 3 times per month for the purposes of detector monitoring. To run the full data reduction chain, we used the \textsc{uves\_obs\_redchain} recipe which performs the following steps: (i) creation of a master bias frame, (ii) implementation of the UVES instrument model, (iii) tracing of Echelle orders positions, (iv)~construction of a master flat field frame, (v)~wavelength calibration, and (vi) reduction of the science frame. The pipeline performs an optimal extraction of the object spectrum and produces the following output files: six spectral segments of spectra with uncertainties (both in merged and resampled formats), and a resampled blaze function. All the output products are one-dimensional. The six segments correspond to different setups, which are produced by the combination of two different optical paths and three CCD detectors. Table~\ref{uves_setups} gives the list of UVES setups and associated wavelength ranges.

\begin{deluxetable}{lc}
\tabletypesize{\footnotesize}
\tablewidth{3pt}
\tablecaption{List of UVES setups and wavelength coverage.\label{uves_setups}}
\tablehead{
\colhead{Dichroic (short name)} & \colhead{Wavelength range, nm} 
}
\startdata 
DIC1 346 blue arm (346B)	& 304 -- 388 \\
DIC2 437 blue arm (437B)	& 373 -- 499 \\
DIC1 580 red arm L (580L)	& 476 -- 577 \\
DIC1 580 red arm U (580U)	& 584 -- 684 \\
DIC2 860 red arm L (860L)	& 660 -- 854 \\
DIC2 860 red arm U (860U)	& 866 -- 1040\\
\enddata
\vspace{-0.5cm}
\end{deluxetable}

\subsection{Calibration and merging of Echelle orders}
\label{subsec:ripples}

\begin{figure}[]
\centering
\includegraphics[width=1\linewidth]{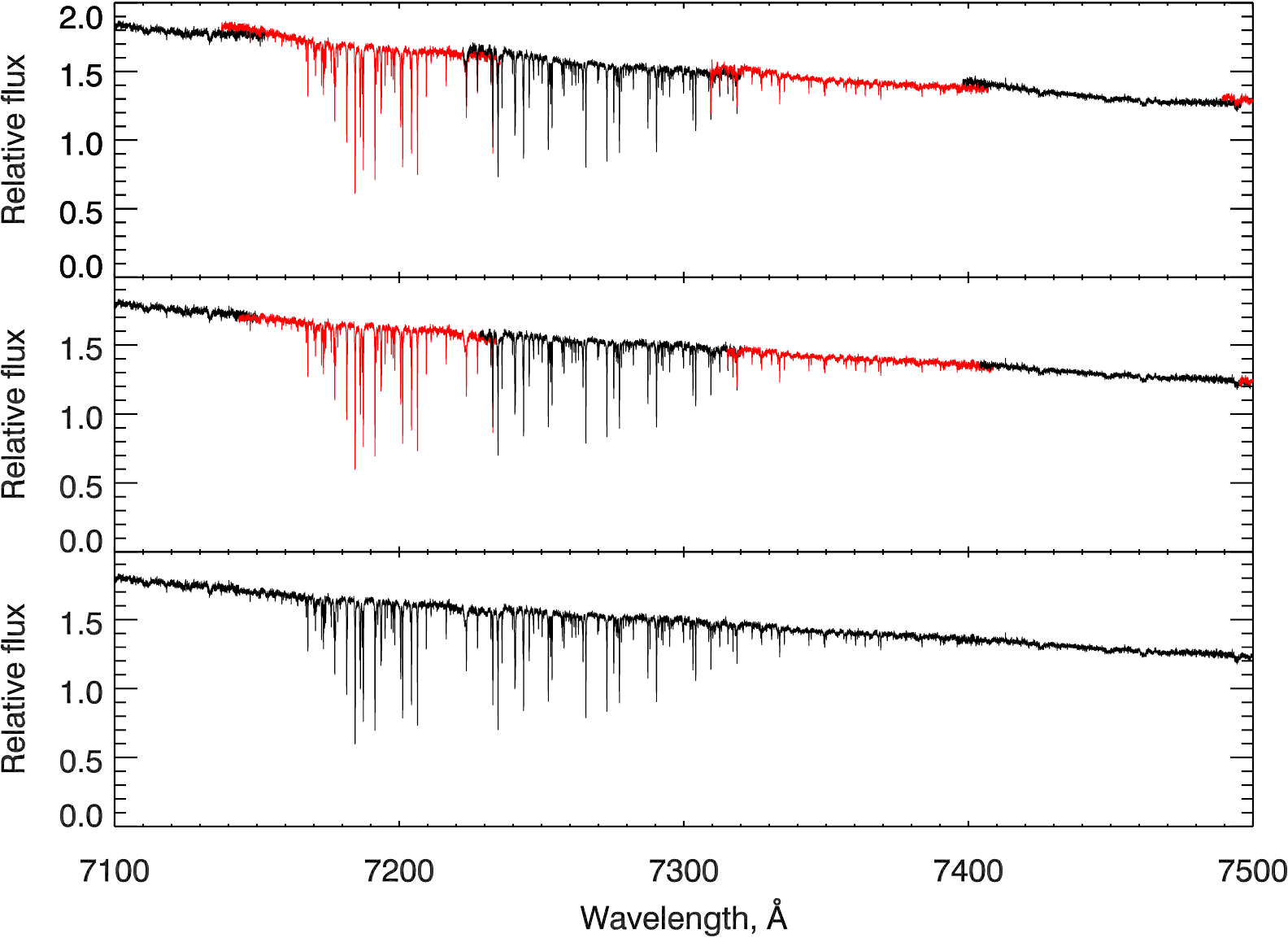}
\caption{The process of Echelle order merging. {\bf Top:} raw orders before applying corrections for blaze shift and scattered light. {\bf Middle:} orders after correction. Orders are highlighted in black or red for visualisation purposes. {\bf Bottom:} spectrum after merging of the corrected orders.}
\label{ord_merging}
\end{figure}

A noticeable feature of the standard pipeline-reduced UVES-POP spectra is a set of ripples in the regions where Echelle orders overlap. In some spectra from the original stellar library,
these ripples substantially deteriorate the final product quality. This occurs for two main reasons. Firstly, the pupil illumination is slightly different between calibrations and science frames.
This results in a different blaze definition and in particular a shift between, e.g., science frames and the internal flat fields \citep{stis_blaze} which are used to correct for the blaze function.
Non-ideal centering of the field derotator, for instance, could cause tiny pupil offsets.
In addition, UVES is not fully sealed so small temperature and/or pressure variations inside the instrument can induce an offset in the wavelength calibration.
The typical offset in UVES is $1/20^{\rm th}$ of a pixel \citep{dekker00} but it can be larger than that in some cases.
The second main possible reason for the existence of ripples in the final spectra is the inaccurate estimation of scattered light and/or of the sky background. Indeed, many observations in UVES-POP were carried out in twilight.
Any over- or under-subtraction of light will lead to the artificial ``bending'' of spectral orders in flux. The top panel of Fig.~\ref{ord_merging} illustrates how spectral orders look as a result of both effects.

We developed an algorithm that allows us to get rid of ripples or significantly reduce their amplitude. The basis of the algorithm is a nonlinear minimization of $\chi^2$, computed as:
\begin{equation}
\chi^2=\sum_{i}^{}\frac{(F^l_i-F^r_i)^2}{\sigma^{_{l^2}}_i+\sigma^{_{r^2}}_i},
\end{equation}
where $F^{l,r}$ and $\sigma^{l,r}$ are fluxes and their errors from the ``left'' and ``right'' overlapping orders respectively. The algorithm finds the values of the shift and over- or under-subtracted scattered light $\Delta_{sl}$. As initial guess, we assumed that $\Delta_{sl}$ is proportional to the mean value of flux in the two adjacent orders, which allows us to take into account scattered light. Our tests have demonstrated that in most cases, scattered light is over-subtracted. After the determination of both values, we corrected the blaze for both of the previously mentioned effects.

\begin{figure}
\centering
\hspace{0.083\hsize}
\includegraphics[width=0.90\hsize]{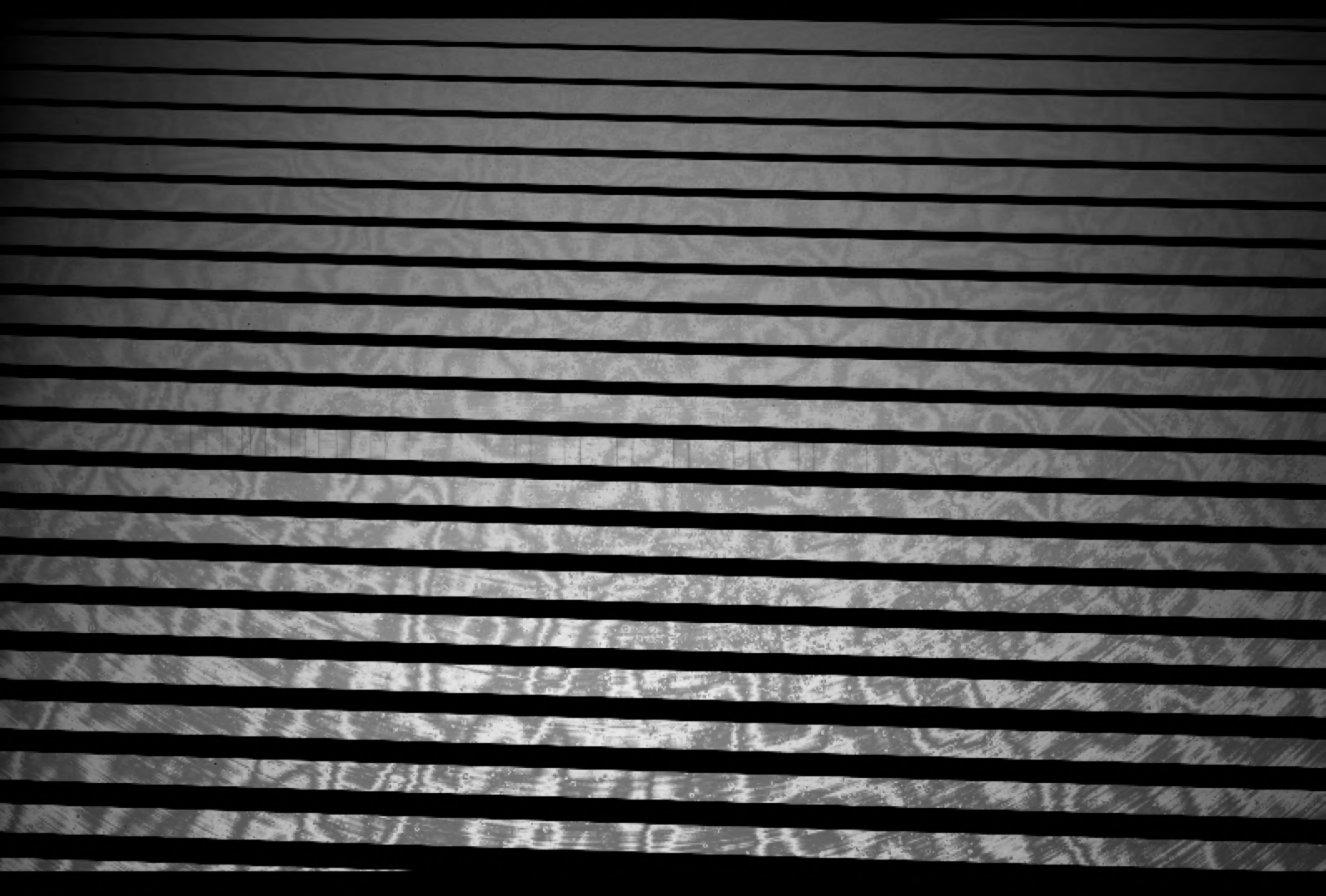}\\
\vskip 1mm
\includegraphics[width=1\hsize]{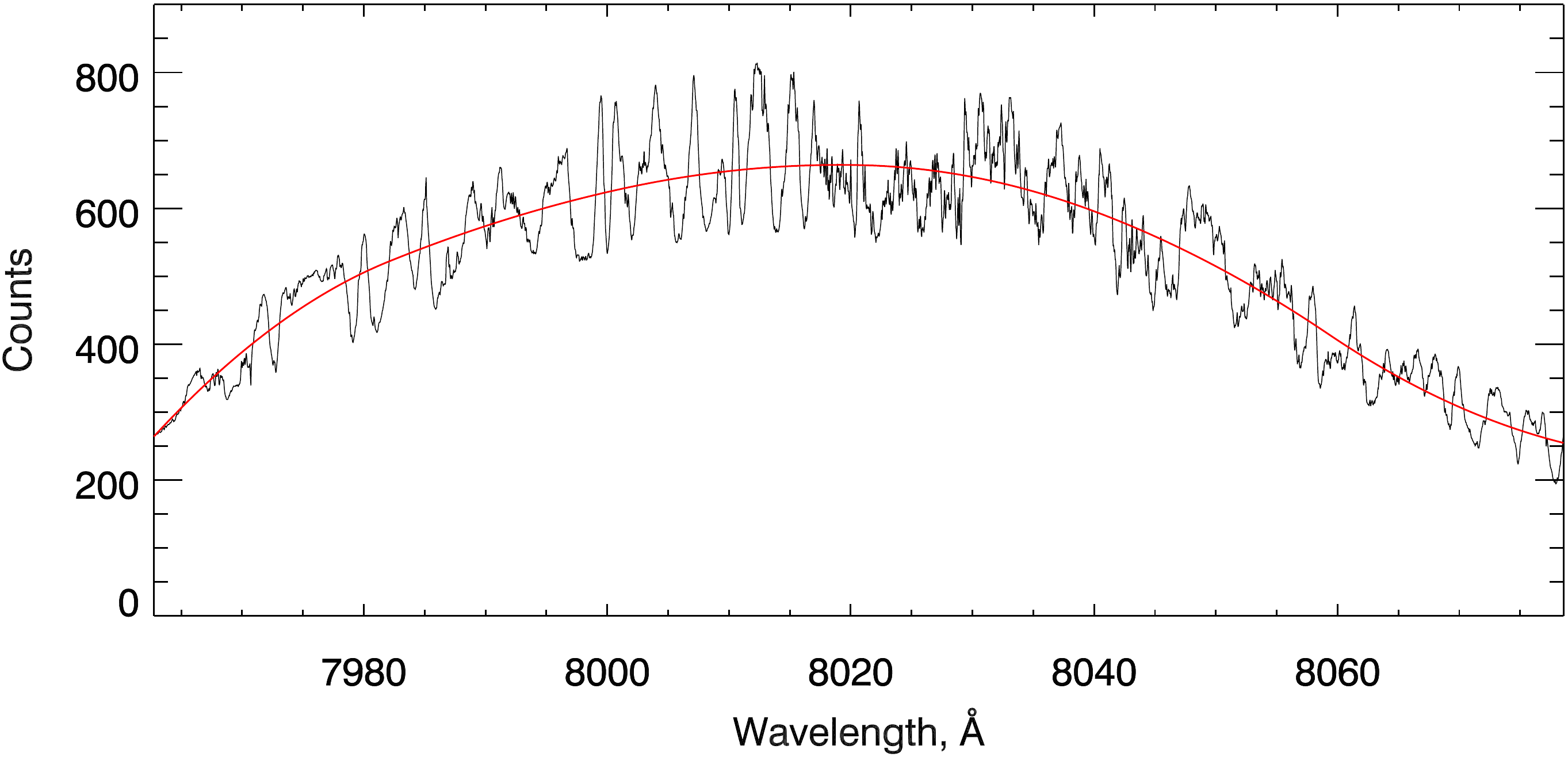}
\caption{{\bf Top:} raw UVES spectral flat field in the 860L segment with strong fringes. {\bf Bottom:} a single flat field order with fringes. The red line shows the best-fitting B-spline which we considered as blaze function approximation.}
\label{fringes}
\end{figure}

The blaze shift procedure is further complicated by the presence of fringes at wavelengths $\lambda \gtrsim 6700$~\AA. Upper panel of Fig.~\ref{fringes} shows fringes on a raw frame. The fact that fringes are strictly fixed in position makes the problem even more complicated. We solved it by decomposing the blaze into two components: the ``real blaze'' component and a fringe pattern (bottom panel of Fig.~\ref{fringes}. We defined the real blaze component as a B-spline approximation to the observed blaze, whereas the ratio of the blaze to this B-spline ($A_{fringes}=F_{blaze}/F_{bspline}$) shows the position and the amplitude of the fringes. The B-spline was shifted by the computed value and then multiplied by $A_{fringes}$ in order to restore the fringe pattern.

With all the orders corrected, we proceeded to merge them into one spectral segment. In the overlapping regions, the flux was averaged with weights proportional to the value of $1/\sigma^2$ (flux uncertainty) and normalized to unity. Because of that, the SNR in these regions is approximately $\sqrt{2}$ times higher. The process of ripple correction and order merging is shown in Fig.~\ref{ord_merging}. 

The described ripple correction algorithm cannot be used in the reddest (860U) segment because Echelle orders do not overlap. For this reason, the correction in this segment was carried out manually by the optimal selection of parameter values. There was no degeneracy of the two parameters (i.e., the value of the blaze shift and the value of over- or underestimation of scattered light) because the blaze shift and inaccurate subtraction of scattered light affect the orders differently: a blaze shift ``tilts'' an order, while inaccurate subtraction of scattered light ``bends'' it.

\subsection{Telluric correction}
\label{subsec:tellcorr}

Observations made with ground-based telescopes suffer from absorption through the Earth's atmosphere which leads to telluric lines imprinted on astronomical spectra. In the wavelength range of UVES, there are absorption bands that mostly originate from water vapor and molecular oxygen. There are also two wide ozone bands. Correction of the spectra for telluric absorption is crucial during data processing.

Telluric correction requires accurate wavelength calibration. It means that the positions of telluric lines and other features in the spectra must precisely match the theoretically computed model telluric lines. For this reason, the barycentric and topocentric corrections were here performed after the telluric correction because telluric features should not be shifted by radial velocity effects.

To perform the telluric correction we need to know how much flux was absorbed on its way through Earth's atmosphere. Spectra can be corrected by dividing the original stellar spectrum by a transmission curve which can be obtained via one of two methods. The first method is to solve the radiative transfer equation of Earth's atmosphere numerically. The second one is to extract the transmission curve based on the spectrum of a star which has as few spectral features as possible. Usually, fast rotating A0V stars are observed during the night. However, in our case, there are no such observations so we chose to calculate the transmission curve. 

We used the Cerro Paranal Advanced Sky Model by ESO (SkyCalc) \citep{skycalc1, skycalc2} to produce a grid of Earth's atmosphere transmission curves in the parameter space of airmass ranging from 1.0 to 2.5 with a fixed step of 0.1 and precipitable water vapor (PWV) in the range 0.5 to 20~mm with increasingly large steps. Our telluric correction procedure approximates an observed stellar spectrum by a linear combination of synthetic stellar templates from the PHOENIX \citep{phoenix} and BT-Settl \citep{bt-settl} libraries using the full spectrum fitting technique. The templates are shifted by the radial velocity amount $v_{rad}$, broadened according to a projected rotational velocity $v\sin i$ and multiplied by a grid of Earth's atmospheric models parametrized by airmass, PWV and then convolved with the spectrograph's line spread function (LSF). We use a multiplicative continuum to account for the global difference between flux-calibrated models and uncalibrated stellar spectra. The airmass, PWV, $v_{rad}$, $v\sin i$, and LSF parameters are fitted non-linearly; the multiplicative polynomial continuum parameters and the weight of the stellar templates in the linear combinations are minimized linearly at each step of the non-linear minimization. Then, telluric absorption was removed by dividing the original spectrum by the telluric model corresponding to the computed values of airmass and PWV and convolved with the LSF. Since the strongest telluric bands are located in the 860L and 860U segments, airmass and PWV values were computed by fitting these segments and these values were then fixed when fitting the remaining segments. Note that the advantage of this telluric correction method is that it can also be applied for fitting galaxy and star cluster spectra.

\begin{figure}
\centering
\includegraphics[width=1\hsize,trim={0 5mm 0 0},clip]{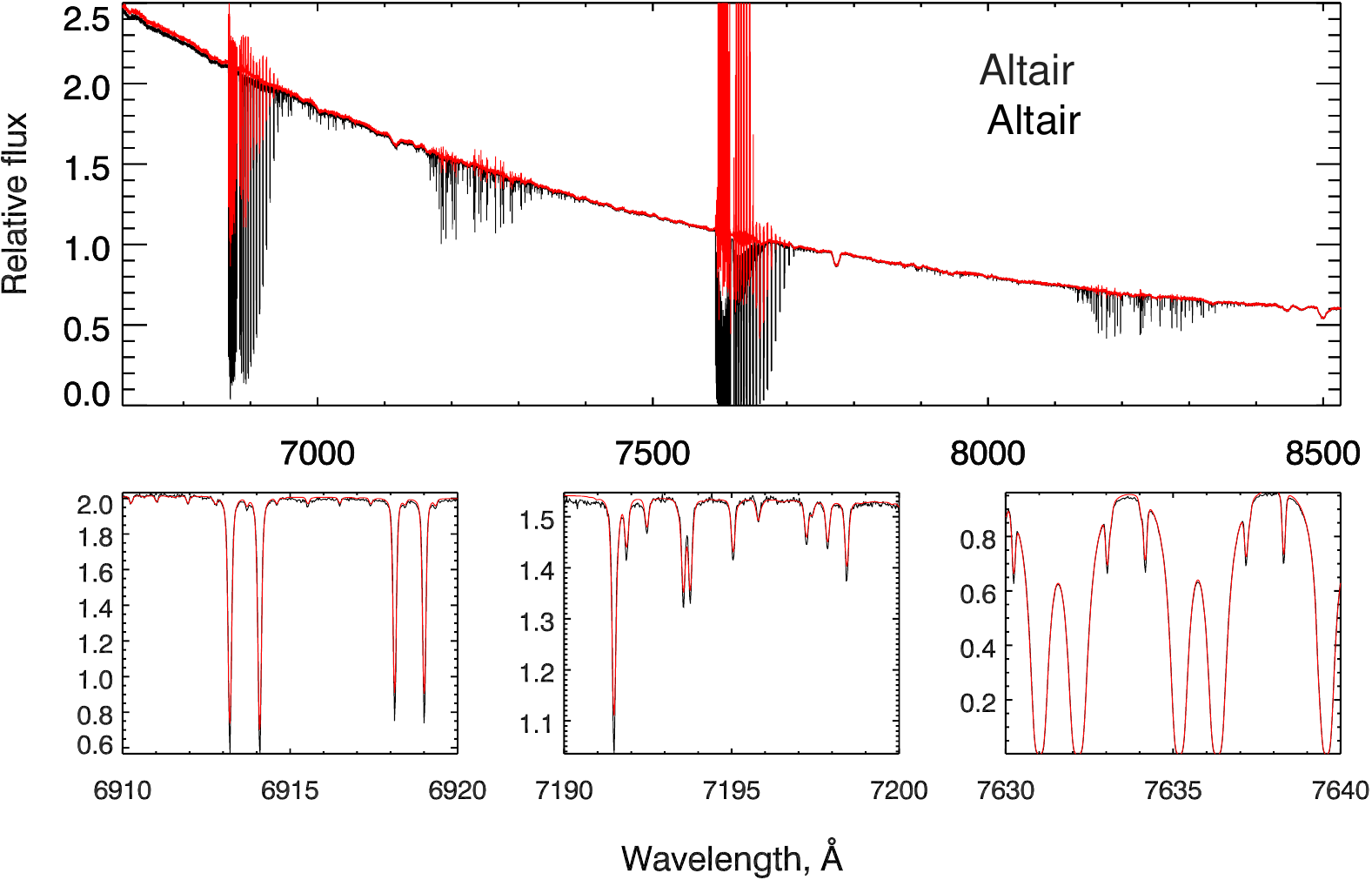}
\vskip 2mm
\includegraphics[width=1\hsize,trim={0 5mm 0 0},clip]{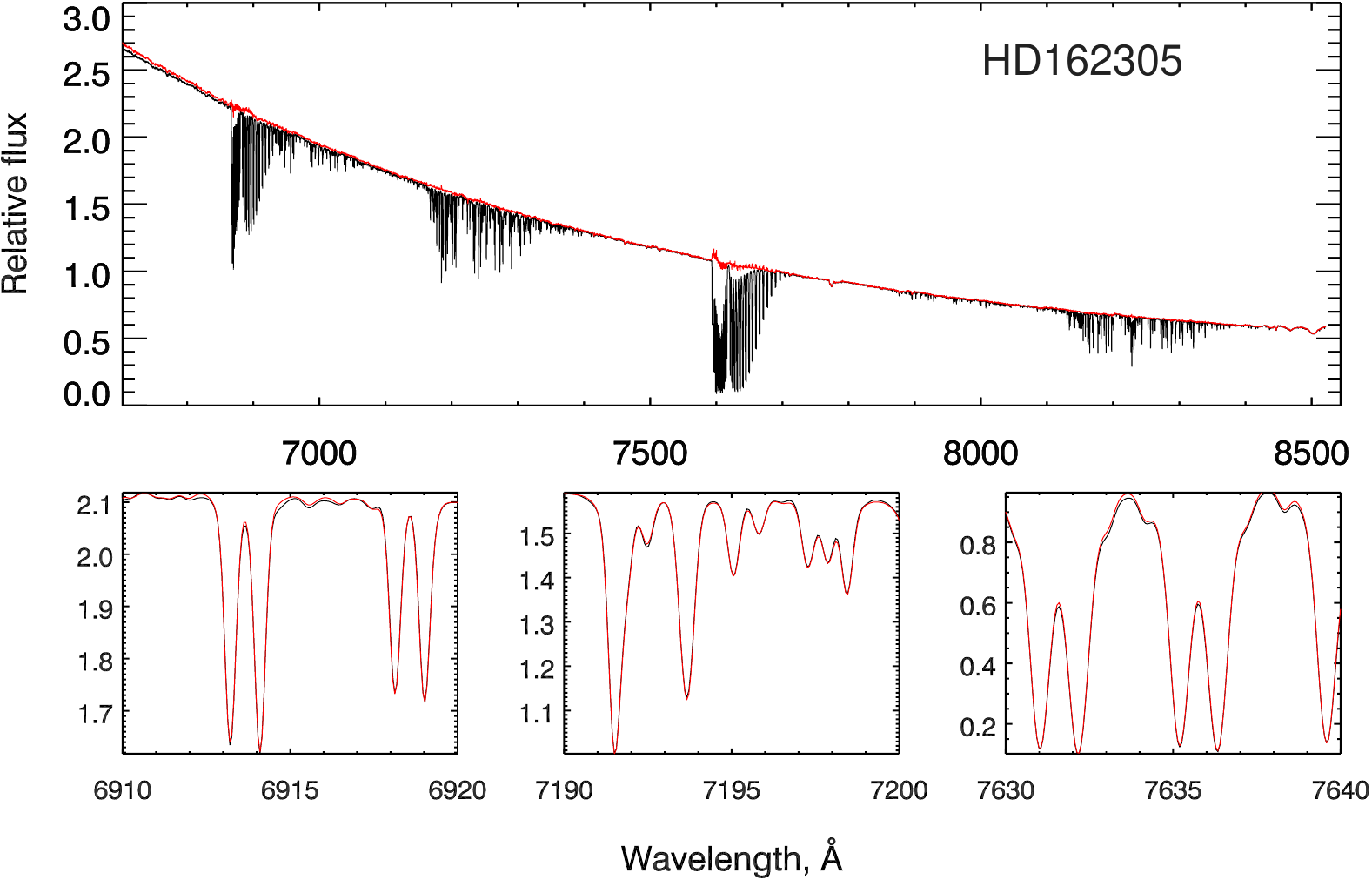}
\vskip 2mm
\includegraphics[width=1\hsize]{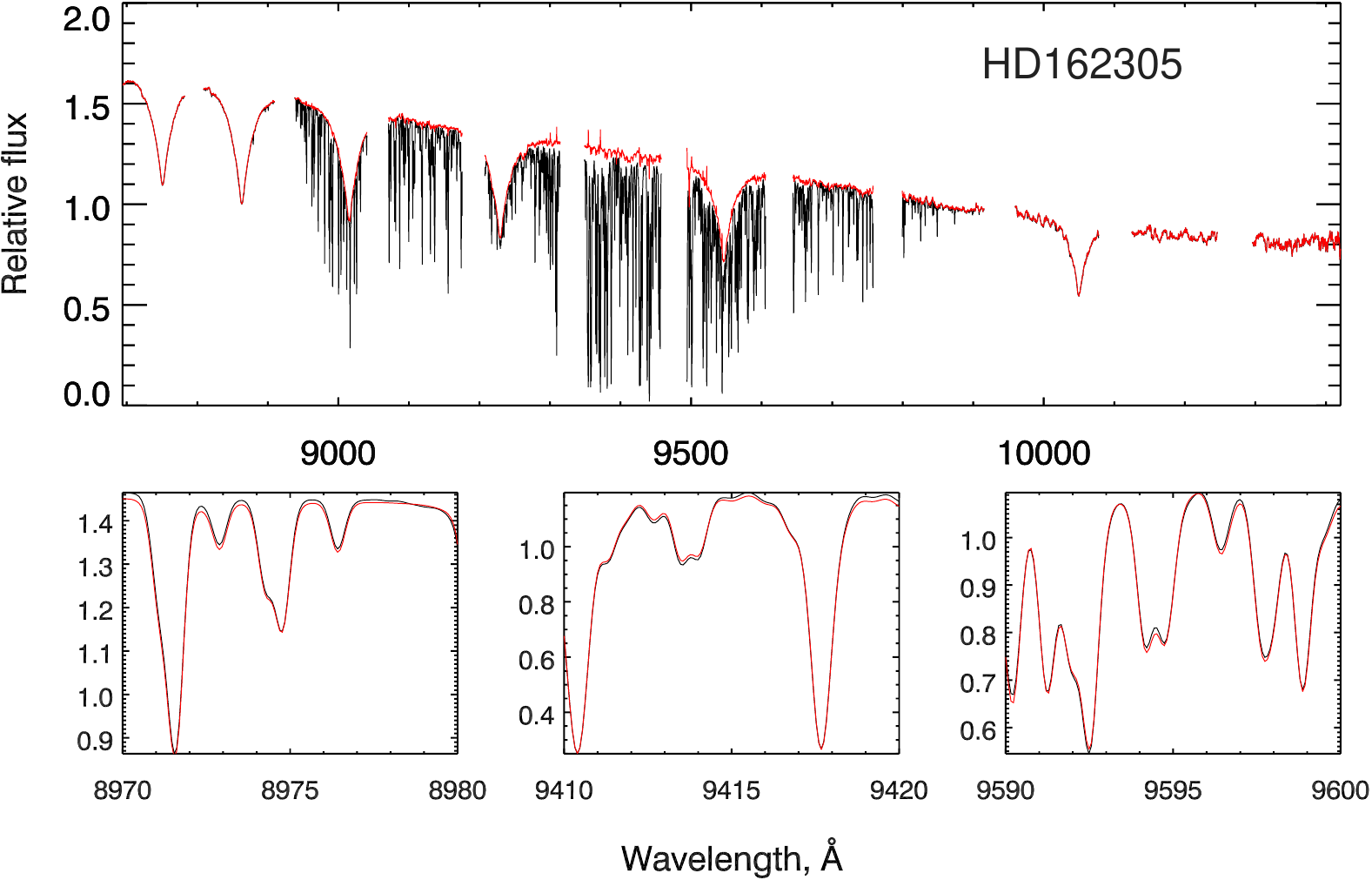}
\caption{{\bf Top:} full-resolution (R=80,000) spectrum of Altair (the 860L segment) before (black) and after (red) the telluric correction. Three smaller panels show the spectrum in three different spectral ranges with the bestfit Earth's atmospheric model (red). {\bf Middle and bottom:} the same as in the upper panels, but for HD~162305 with R=20,000.}
\label{tellur_ex}
\end{figure}

Unfortunately, the quality of telluric correction is rather poor at the full spectral resolution. It might be caused by either improper centering of a star on the slit during observation or an imperfect correction of the atmospheric dispersion. Another factor might be the insufficient quality of the Earth's atmosphere transmission models (e.g., incomplete molecular line lists or transition probabilities) leading to similar effects, however, there have been no serious issues reported in the literature when using SkyCalc on full-resolution UVES spectra obtained through a fiber input. All UVES-POP stars were observed through a slit, therefore the atmospheric seeing quality could have strong effect on the final spectral resolution if the slit was too wide (which was not often the case because of the narrow 0.5$''$-wide slit used in the observations). But even subtle seeing variations would cause the effective spectral resolution change and lead to artefacts in the telluric correction. Because of this, the quality of telluric correction of the spectra with the original resolving power R=80,000 is often unsatisfactory in the regions of high absorption (see Fig.~\ref{tellur_ex} top panel).

At the same time, at the reduced resolving power of R=20,000 the telluric correction quality is  excellent, and this spectral resolution is sufficient for most applications. It fully covers the needs of stellar population synthesis: R=20,000 corresponds to the instrumental resolution of $\sim$6.4~km s$^{-1}$, which is significantly smaller than the typical velocity dispersion of galaxies or most globular clusters. Fig.~\ref{tellur_ex} middle and bottom panels demonstrate the telluric correction for HD~162305 in two different spectral segments. Although the quality of the telluric correction at the reduced resolution is good, there are some artifacts because of imperfect atmospheric transmission models, the most noticeable artifact is at $\sim$7600~\AA looking like a small bump. 

The main release of our recalibrated dataset presented here and available via interactive web-based visualisation tools was obtained by convolving the original R=80,000 spectra with a Gaussian LSF corresponding to R=20,000. From now on in this manuscript, we will be dealing with the spectra having a reduced resolving power of R=20,000 unless noted otherwise. We also provide full-resolution spectra for download through the web-site and Virtual Observatory access mechanisms (see Section~\ref{sec:web}). 

\subsection{Obtaining UVES response curves}
\label{subsec:respcurv}

After performing the telluric correction, spectra need to be flux calibrated. Since the original spectra taken with UVES are not ripple- and telluric corrected, the original response curves\footnote{\url{http://www.eso.org/observing/dfo/quality/UVES/qc/SysEffic_qc1.html##response}} which were derived based on these spectra \citep[see][]{uves_rc} sometimes have poor quality. We decided to re-compute the response curves using the corrected spectra.

As a preparation for this stage, we performed a preliminary determination of stellar atmospheric parameters (see Section~\ref{sec:atmpar}) using spectra corrected with original response curves which were accurate enough for this purpose. The response curves changed several times over the entire period of observation of UVES-POP due to mirror re-coating or cleaning and filters replacement. Thus, we need to derive response curves for each period of time, i.e, the time interval between two successive changes in the optics of the telescope or the spectrograph. The dates of these changes are documented and presented on the web page of the European Southern Observatory. For each period, we selected stars that, firstly, have a relatively small number of spectral lines and, secondly, the fitting quality of which was very good. 

For the selected stars, we obtained model spectra corresponding to the atmospheric parameters of these stars. Model spectra were artificially ``reddened'' by the value of color excess  $E(B-V)$ taken from the literature using the extinction law from \citet{Fitzpatrick1999}. The response curves were defined as B-splines which fit the ratios of model spectra to the observed ones. Parts of spectra, which could not be adequately described by models were masked out.

\subsection{Merging of UVES spectral segments}
\label{subsec:merging}

A full spectrum obtained with UVES consists of 6 segments, some of which overlap. There are also two gaps, one between the segments 580L and 580U and another one between the segments 860L and 860U. Three segments: 346B, 437B, and 580L, and also the two segments 580U and 860L, can be merged into two long segments of 255~nm and 268~nm respectively. Thus, we used two different merging techniques, one of which is used to stitch overlapping segments and another one for non-overlapping ones.

\begin{figure}
\centering
\includegraphics[width=1\linewidth]{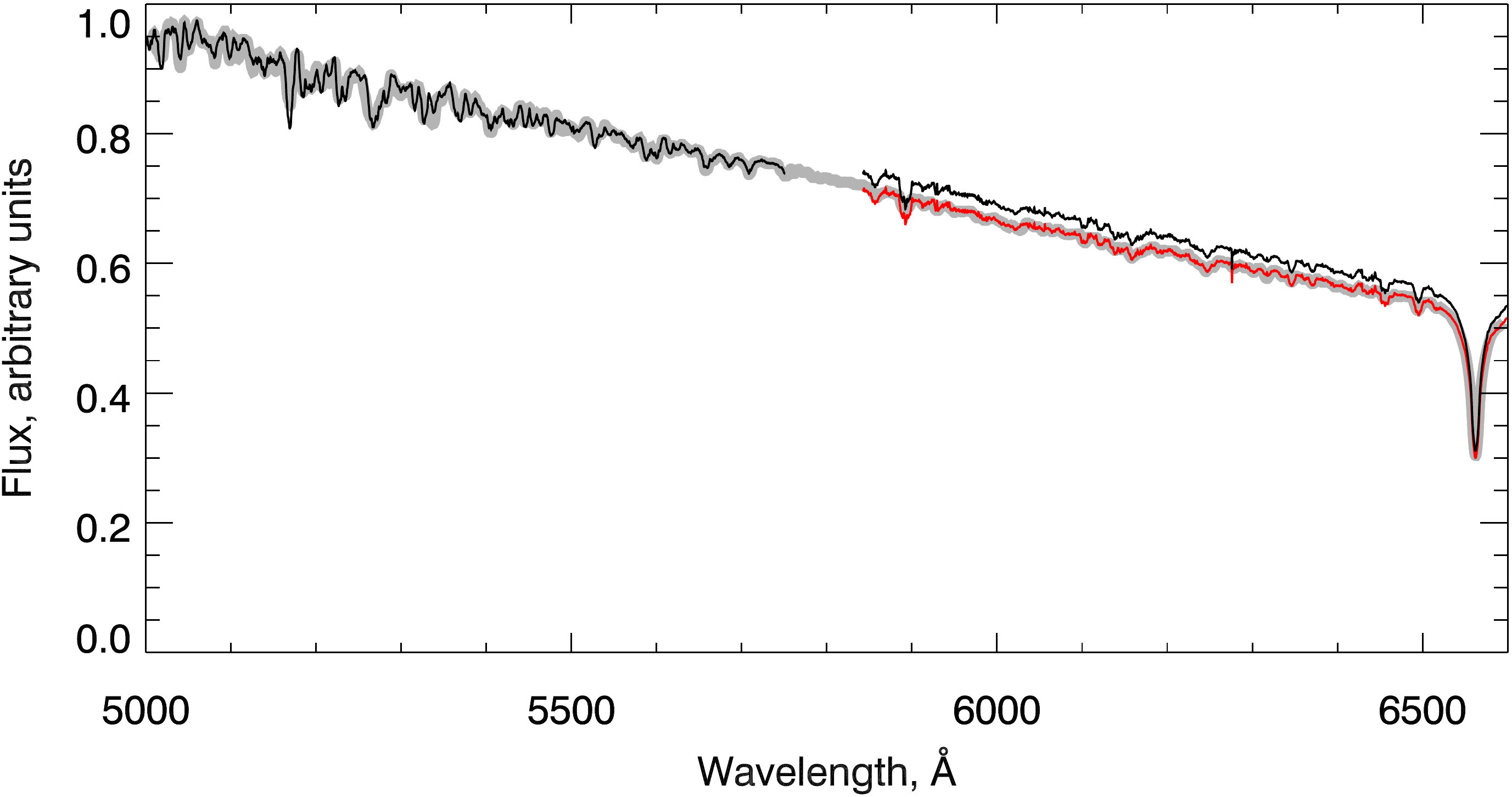}
\caption{Merging of non-overlapping spectral segments for Altair. The black line shows the spectra prior to merging. The red line shows the right segment re-normalized by the merging procedure. A synthetic spectrum used for the merging of non-overlapping segments is shown in grey.}
\label{merge_non}
\end{figure}

In the case of overlapping segments, we calculated the flux in the overlapping regions as the weighted mean of the fluxes of the two segments. Weights were determined as coefficients that linearly increase from 0 at the edge of a segment to 1 within the overlapping region. To merge non-overlapping segments, we developed an algorithm using synthetic spectra (PHOENIX; \citealp{phoenix}, and BT-Settl; \citealp{bt-settl}) and stellar atmospheric parameters which we obtained by fitting one of the segments (see section~\ref{sec:atmpar}). This allows us to estimate the flux difference on both sides of each gap and avoid ``steps'' in flux in the final spectrum. Fig.~\ref{merge_non} shows an example of the merging of the non-overlapping segments.

Before merging, we corrected all spectral segments for Doppler shift caused by barycentric and topocentric velocities. The maximum absolute value of topocentric velocity at the latitude of Cerro Paranal is 423~m s$^{-1}$ which is especially significant for high-resolution spectra since this value is around 27\%\ of the FWHM at the resolving power of R=80,000.

\subsection{Variable stars}
\label{subsec:var_stars}

Our sample contains 106 variable stars of various variability types: 51 pulsating variables, 14 eruptive variables, 27 rotating variables, 4 eclipsing binaries, and 10 stars of other types. The physics of variability causes changes in the observed radial velocity in the case of pulsating and eclipsing binaries, as well as surface gravity, and effective temperature due to stellar pulsations.

We obtained the variability types for these stars and their periods (where applicable and available) from the General Catalogue of Variable Stars \citep{gcvs}. For some stars, the periods are not available, however, their published variability type puts limitations on possible period values. We compared the periods of the variable stars of the sample with the differences of starting time between spectral exposures. In all the checked cases for pulsating variables, this difference was much smaller than the period (or typical period) of the corresponding variable stars. Therefore, the different spectral setups were observed in nearly the same variability phase. Thus, the segment merging did not require velocity corrections for individual segments and ensured that the entire spectrum would correspond to the same stellar atmospheric parameters.

We provide the information on stellar variability in the main data table presented in Appendix~B and available electronically alongside this manuscript.

\subsection{Spectrophotometric correction}
\label{subsec:phot_sec}

Because of varying observing conditions during the night and imperfections of synthetic stellar atmospheres, the response curves obtained as described above when applied to observed spectra of other stars, can produce systematic flux calibration uncertainties of up to 20--30~\%. Therefore, we performed an additional spectrophotometric correction of the final spectra by using published photometric measurements of the corresponding stars. At the end, we brought the flux scale in the UVES spectra to physical units erg~s$^{-1}$~cm$^{-2}$~\AA$^{-1}$.

To perform the spectrophotometric correction, we used the broad- and middle-band photometric data from the catalogs \textit{Gaia}~DR2 \citep{Prusti2016,gaia_hr,gaia_bands}, Tycho-2 \citep{tycho_cat,tycho_bands}, WBVR catalog \citep{almata}, 13-color photometric catalog \citep{color13}, APASS \citep{apass}, Cousins catalog \citep{cousins}, the Vilnius catalog \citep{vilnius}, as well as magnitudes in the $Z$-VISTA and $Y$-VISTA bands \citep{vista2}, recalculated from $J$ and $K_s$ 2MASS magnitudes \citep{mass2} using formulae from \citet{mass2vista}. The transmission curves of all the photometric filters used in the correction procedure are shown in the Fig.~\ref{bands}. 

First, we converted the data from all catalogs into the Vega system. This was done for the data in the $g$, $r$, and $i$ bands of the APASS catalog which uses the $AB$~system \citep{ab_sys}, and data from the Vilnius catalog which has its own specific photometric system. The magnitudes in the $g$, $r$, and $i$ were converted to the Vega system using zero-point corrections from ~\citet{sdss2vega}. We performed the zero-point conversion of the Vilnius data to the Vega system by comparing the Vega magnitudes in this catalog with the value of $0.03^m$ \citep{Bessel2005}. Also, we corrected \textit{Gaia} magnitudes of bright stars which suffered from saturation according to \citet{gaia_corr}.

To ensure the high quality of the spectrophotometric correction, we need to use the photometric bands with the effective wavelengths covering the entire observed spectral range. It is especially important to use the bands with $\lambda_{\mathrm{eff}}$ close to the edges of the UVES spectral range. Because most red filters with $\lambda_{\mathrm{eff}}>9000$~\AA~have transmission curves which extend beyond the spectral range of UVES, we extrapolated the spectra using best-fitting synthetic stellar atmospheres for the corresponding spectra. This procedure was necessary to use the data in the band \textit{Gaia}~R$_\mathrm{p}$ ($\lambda_{\mathrm{eff}}=7418$~\AA$<9000$~\AA), $Y$-VISTA ($\lambda_{\mathrm{eff}}=10184$~\AA) and the `99' band of the 13-color catalog ($\lambda_{\mathrm{eff}}=9818$~\AA). Also, we used the same best-fitting stellar templates to fill the gaps between segments and/or individual orders.

\begin{figure}[]
\centering
\includegraphics[width=1\linewidth]{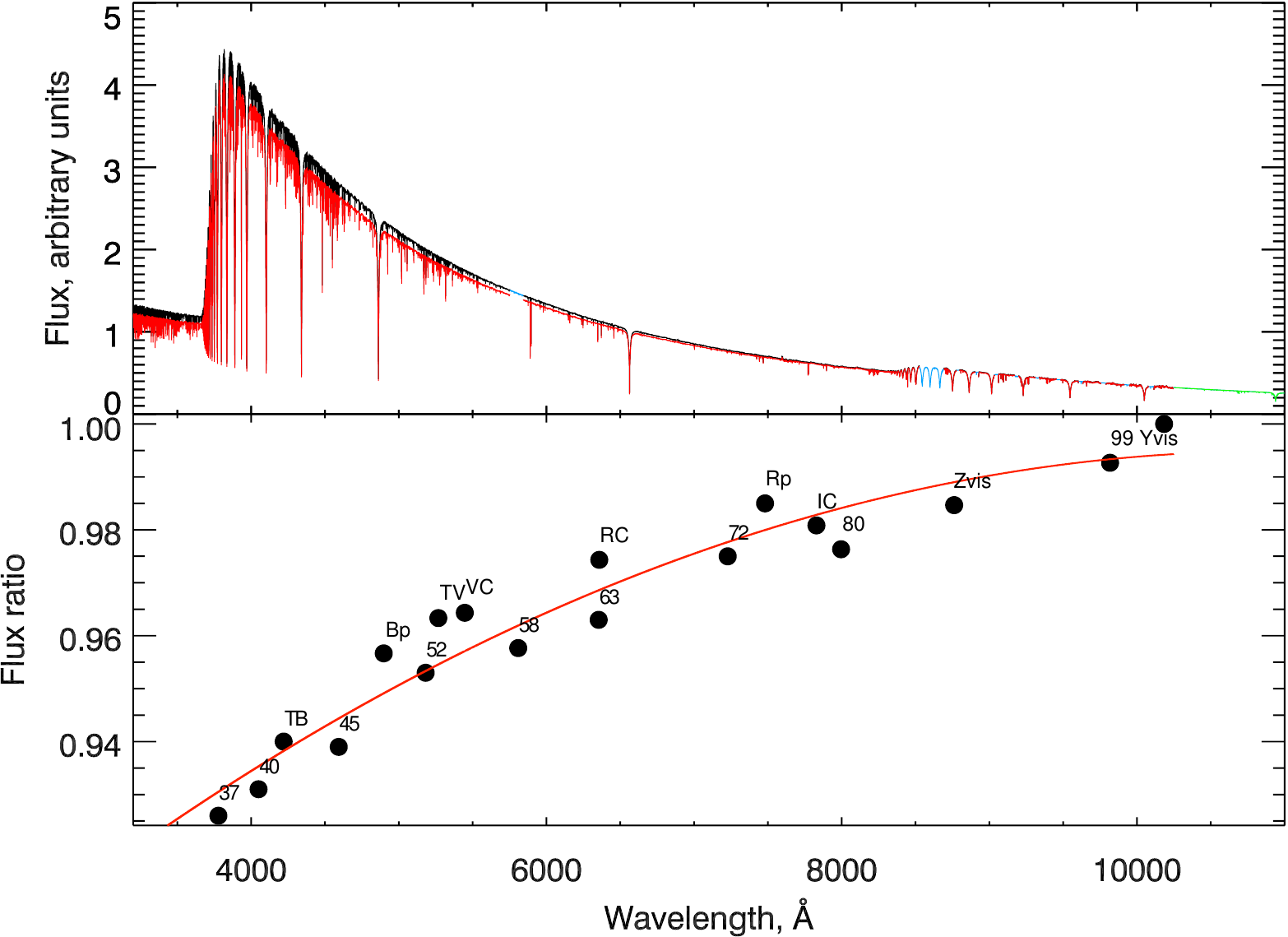}
\caption{Spectrophotometric correction for the UVES-POP spectrum of HD~75063. {\bf Top:} spectrum before correction (black line) and after correction (red line). Blue regions demonstrate the best-fitting synthetic spectrum which fills in the gaps, while the green line is an extrapolated spectrum. {\bf Bottom:} black points are the ratios of observed fluxes from photometric databases and synthetic fluxes computed from the spectrum (the ratios are normalized to the maximal value). The red line is a polynomial fit used for the correction. Explanation of abbreviations and some parameters of the photometric bands are shown in Table~\ref{bands_tab}.}
\label{phot_ex}
\end{figure}

\begin{figure}[]
\centering
\includegraphics[width=1\linewidth]{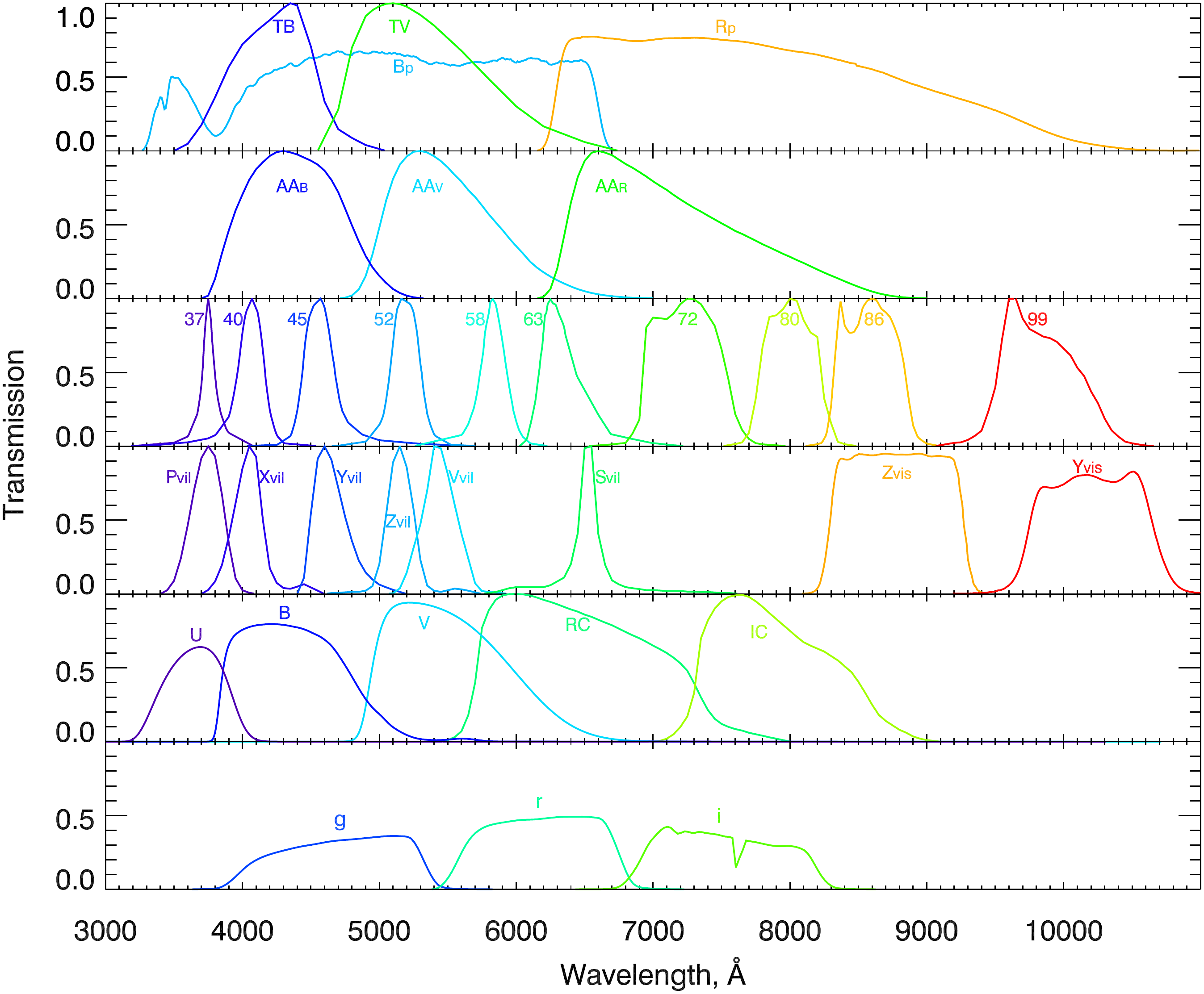}
\caption{Transmission curves of the photometric bands used for the spectrophotometric correction. Explanation of abbreviations and parameters of the bands are presented in Table~\ref{bands_tab}.}
\label{bands}
\end{figure}

The correction algorithm fits a low-order polynomial which transforms an observed spectrum in such a way that the synthetic magnitudes fit the observed ones best. An example of spectrophotometric correction is presented in Fig.~\ref{phot_ex}. The algorithm iteratively excludes some data points by using a rejection technique for outliers, which could occur because of high photometric errors or stellar variability. Also, a wrong cross-match of a star from the UVES-POP list against photometric catalogues could cause these outliers, which is especially relevant for high-proper motion stars. In total, we performed spectrophotometric correction for 306 stars from the library.

\begin{deluxetable}{lcc|lcc}
\tabletypesize{\footnotesize}
\tablewidth{3pt}
\tablecaption{The photometric bands used for spectrophotometric correction and some of their properties. \label{bands_tab}}
\tablehead{
\colhead{Band} & \colhead{Phot.Sys.} & \colhead{$\lambda_\mathrm{eff}$,~\AA} & \colhead{Band} & \colhead{Phot.Sys.} & \colhead{$\lambda_\mathrm{eff}$,~\AA}
}
\startdata 
B$_p$  	    	& \textit{Gaia}    	        & 5044.4 & P$_{vil}$    	& Vilnius    	            & 3768.9 \\
R$_p$ 		    & \textit{Gaia}   	        & 7692.2 & X$_{vil}$    	& Vilnius    	            & 4051.3 \\
TB  		    & Tycho  	            	& 4220.0 & Y$_{vil}$    	& Vilnius    	            & 4649.7 \\
TV 		        & Tycho  	            	& 5266.4 & Z$_{vil}$    	& Vilnius    	            & 5151.9 \\
AA$_B$      	& WBVR    	                & 4347.2 & V$_{vil}$     	& Vilnius    	            & 5432.4 \\
AA$_V$      	& WBVR    	                & 5442.2 & S$_{vil}$    	& Vilnius    	            & 6500.3 \\
AA$_R$      	& WBVR    	                & 7008.8 & Z$_{vis}$    	& VISTA                 	& 8762.4 \\
37  	    	& 13-color                  & 3774.8 & Y$_{vis}$    	& VISTA                 	& 10184.2 \\
40  	    	& 13-color                  & 4046.9 & U            	& Johnson                   & 3663.6 \\
45  	    	& 13-color                  & 4586.3 & B              	& Johnson                   & 4360.0 \\ 
52  	       	& 13-color                  & 5180.2 & V              	& Johnson                   & 5445.8 \\
58  	    	& 13-color                  & 5806.2 & RC           	& Cousins                   & 6358.0 \\
63  	    	& 13-color                  & 6349.1 & IC            	& Cousins                   & 7829.2 \\
72      		& 13-color                  & 7222.7 & g             	& SDSS                      & 4673.0 \\
80      		& 13-color                  & 7993.6 & r             	& SDSS                      & 6142.0 \\
86      		& 13-color                  & 8577.7 & i            	& SDSS                      & 7459.0 \\
99      		& 13-color.                 & 9813.8 &                  &                           &        \\
\enddata
\vspace{-0.5cm}
\end{deluxetable}

\begin{figure}[]
\centering
\includegraphics[width=1\linewidth]{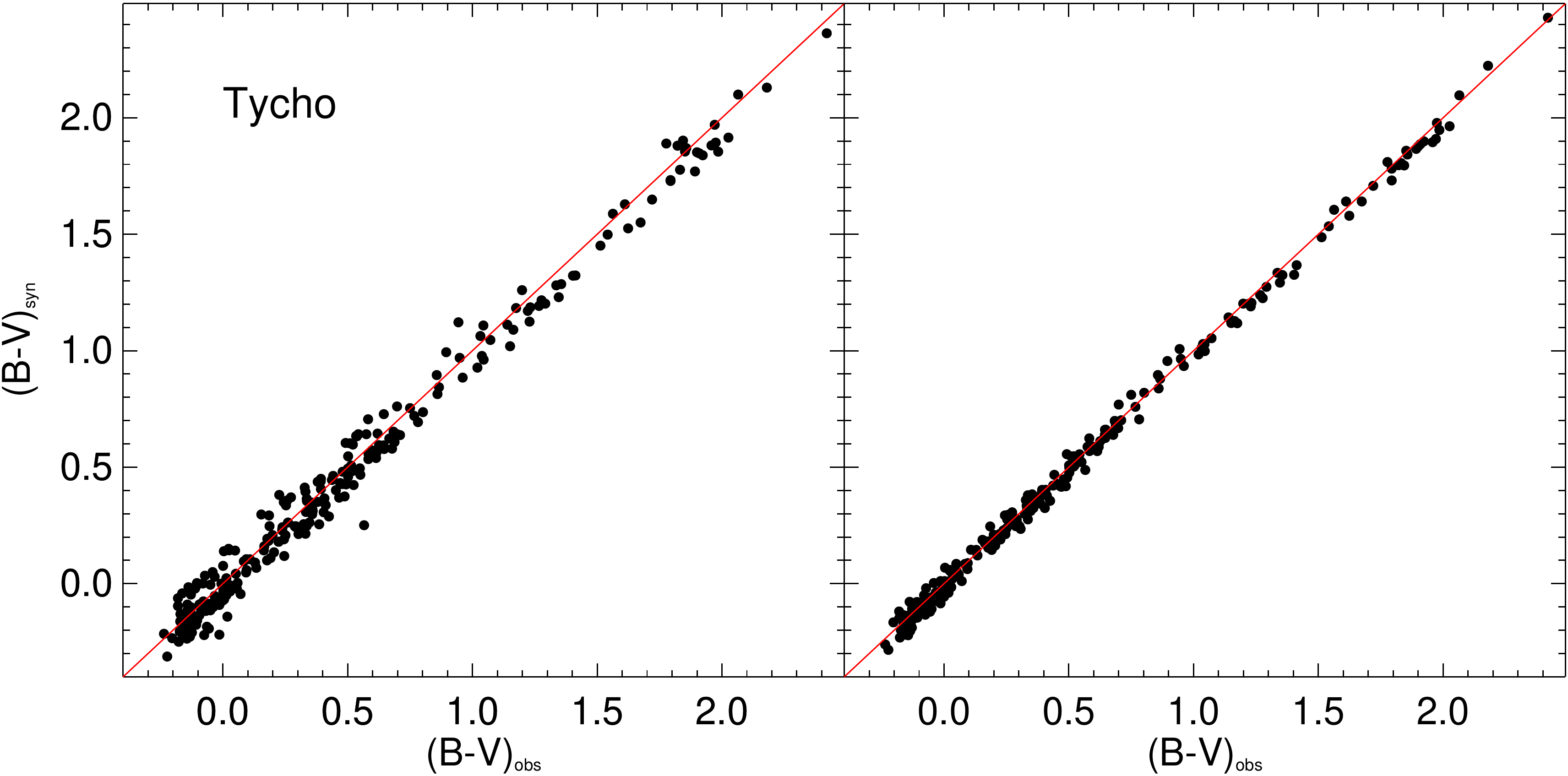}
\includegraphics[width=1\linewidth]{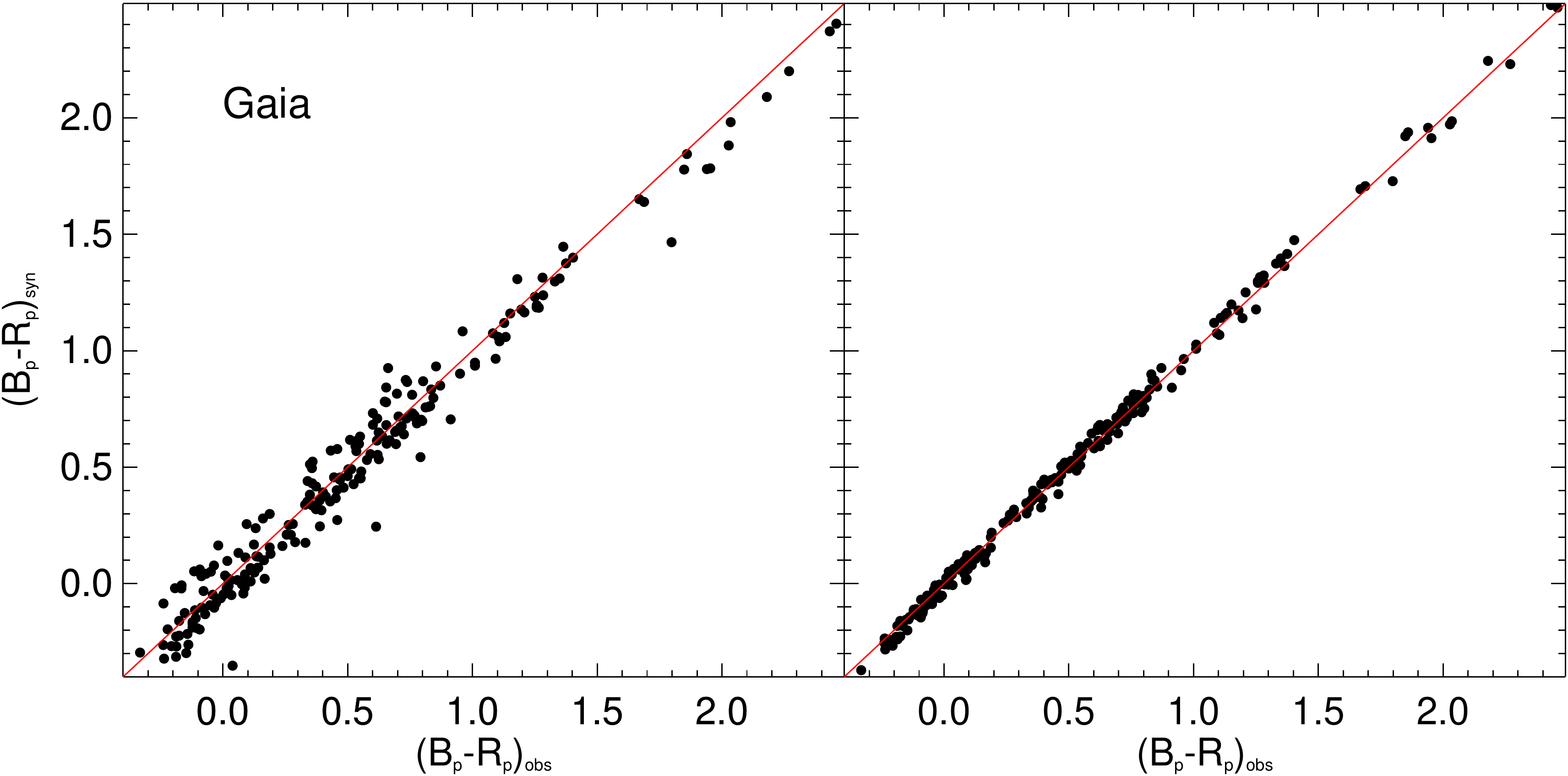}
\caption{{\bf Top:} Comparison of the synthetic Tycho colors $(B-V)_{syn}$ of the spectra corrected with observed colors $(B-V)_{obs}$ before and after spectrophotometric correction (left and right panels, respectively ). {\bf Bottom:} Same as top but for \textit{Gaia} colors $(B_p-R_p)$.}
\label{comp_phot}
\end{figure}

To check the quality of the spectrophotometric correction, we compared the synthetic colors in the Tycho-2 and \textit{Gaia} photometric systems with the observed colors. These catalogs contain data for more stars from the UVES-POP library than any other catalog used. In addition, these observations do not suffer from the effects of the Earth's atmosphere. Comparison of the synthetic colors of the spectra corrected with observed colors is presented in the Fig.~\ref{comp_phot}. Robust standard deviation of the value $(B-V)_{obs}-(B-V)_{syn}$ before the spectrophotometric correction is 0.071~mag and after the correction it is improved to 0.033~mag, for the \textit{Gaia} color $(B_p-R_p)$ the corresponding values are 0.085~mag and 0.034~mag.

\subsection{Interstellar extinction}
\label{subsec:ebv_sec}

Calculating interstellar extinction for sources within the Galaxy either requires the knowledge of the three-dimensional dust distribution known to be strongly non-uniform and notoriously hard to derive, or can be computed using secondary indicators such as interstellar absorption lines of sodium (NaD) or calcium (Ca H\&K), which correlate with the line-of-sight amount of dust \citep[e.g.,][]{Hobbs1974, Poznanski2012}.

We computed the color excess $E(B-V)$ for 41 stars using the publicly available code \texttt{dustmap}\footnote{https://dustmaps.readthedocs.io/en/latest/index.html}. The algorithm is based on the integration of the extinction along the line-of-sight using a 3D map of dust in the Milky Way reconstructed from optical-to-near-infrared photometry of stars \citep{Green2015,Green2018}. For the remaining 365 stars it was not possible to compute the $E(B-V)$ values because the published 3D map does not cover a substantial portion of the Southern hemisphere. We independently determined $E(B-V)$ for 364 stars with the parameters of stellar atmospheres described in Section~\ref{sec:atmpar} by approximating the multiplicative continuum returned by the fitting code with the extinction curve from \citet{Fitzpatrick1999}.

\section{Determination of stellar atmospheric parameters}
\label{sec:atmpar}

\begin{figure*}[]
\centering
\includegraphics[width=\hsize]{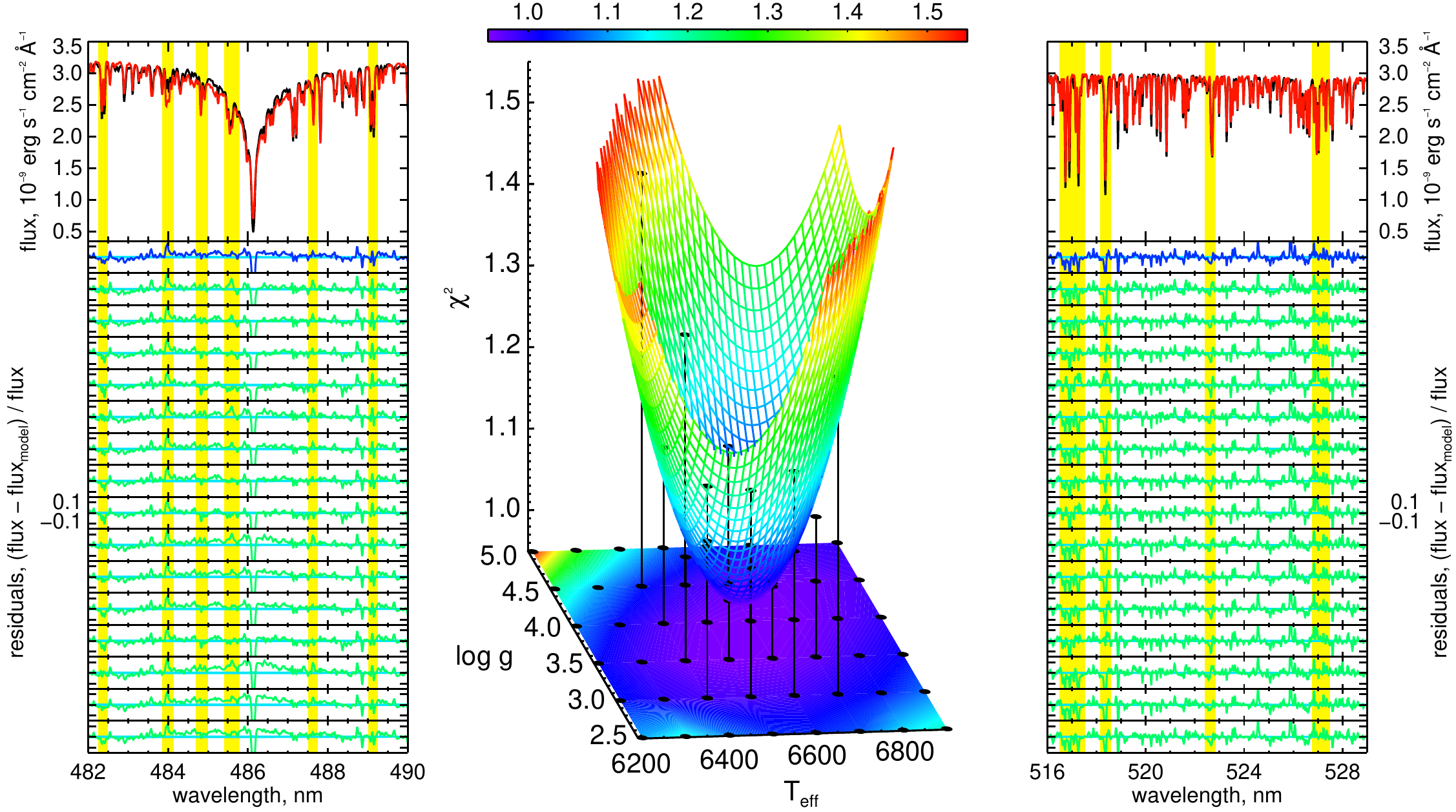}
\caption{
Visualization of the final stage of the minimization method using for the fitting of a spectrum of Procyon by grids of stellar atmosphere models with the two discrete parameters ($T_{\mathrm{eff}}$, $\mathrm{log}~g$ assuming Solar metallicity and [$\alpha$/Fe]), and continuous kinematic parameters ($v_{rad}$, $v \sin i$) and a multiplicative continuum.
The central panel shows a 3D-plot of the 2-nd order approximation surface for the $\chi^2$ profile.
The projection of the $\chi^2$ profile is shown in the same colors, the black dots mark the grid nodes in which the $\chi^2$ values were calculated.
The position of the nodes selected for the approximation are also shown on a 3D surface.
The right and left panels present two small portions of a spectrum around H$\beta$ and the Mg$b$ triplet, their best-fitting models and fitting residuals in every node marked in the middle panel in the decreasing order of $T_{\mathrm{eff}}$ and $\mathrm{log}~g$. The input data are shown in black, the best-fitting model is shown in red, the residuals are in blue, the residuals in the grid nodes selected for the approximation are shown in green in the same flux scale, the spectral lines with the most noticeable changes from the parameters of stellar atmospheres are highlighted in yellow.}
\label{method_procyon}
\end{figure*}

\begin{figure*}
\includegraphics[width=\textwidth]{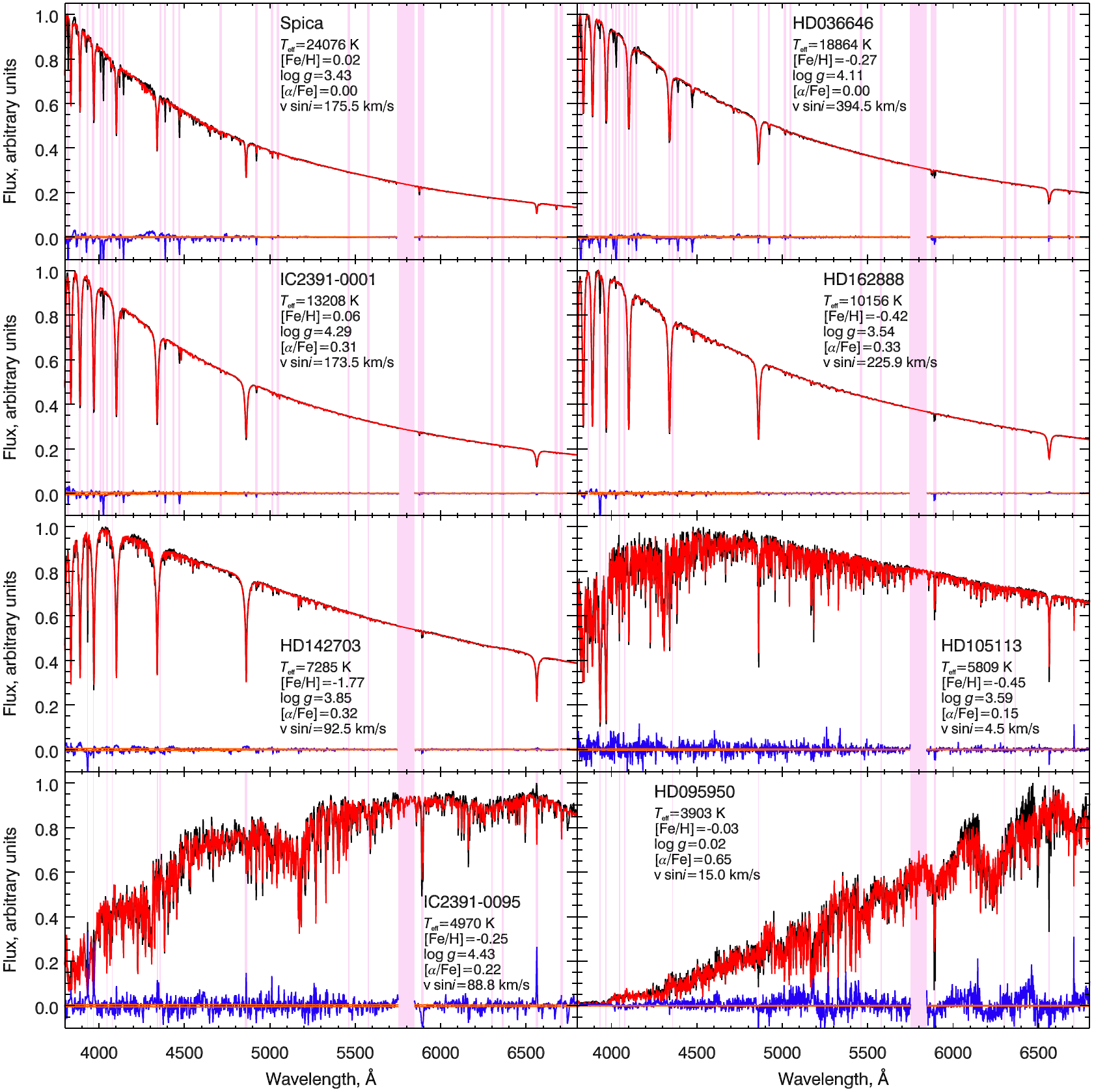}
\caption{Examples of full spectrum fitting of stellar spectra. The observed spectra are shown in black in the fitting range from 3800--6800~\AA, the uncertainties are in orange, the best-fitting models are in red, and the fitting residuals are in blue. Pink-shaded areas denote the masked regions excluded from the fit.}
\label{fit_examples}
\end{figure*}

\begin{figure}[]
\centering
\includegraphics[width=1\hsize]{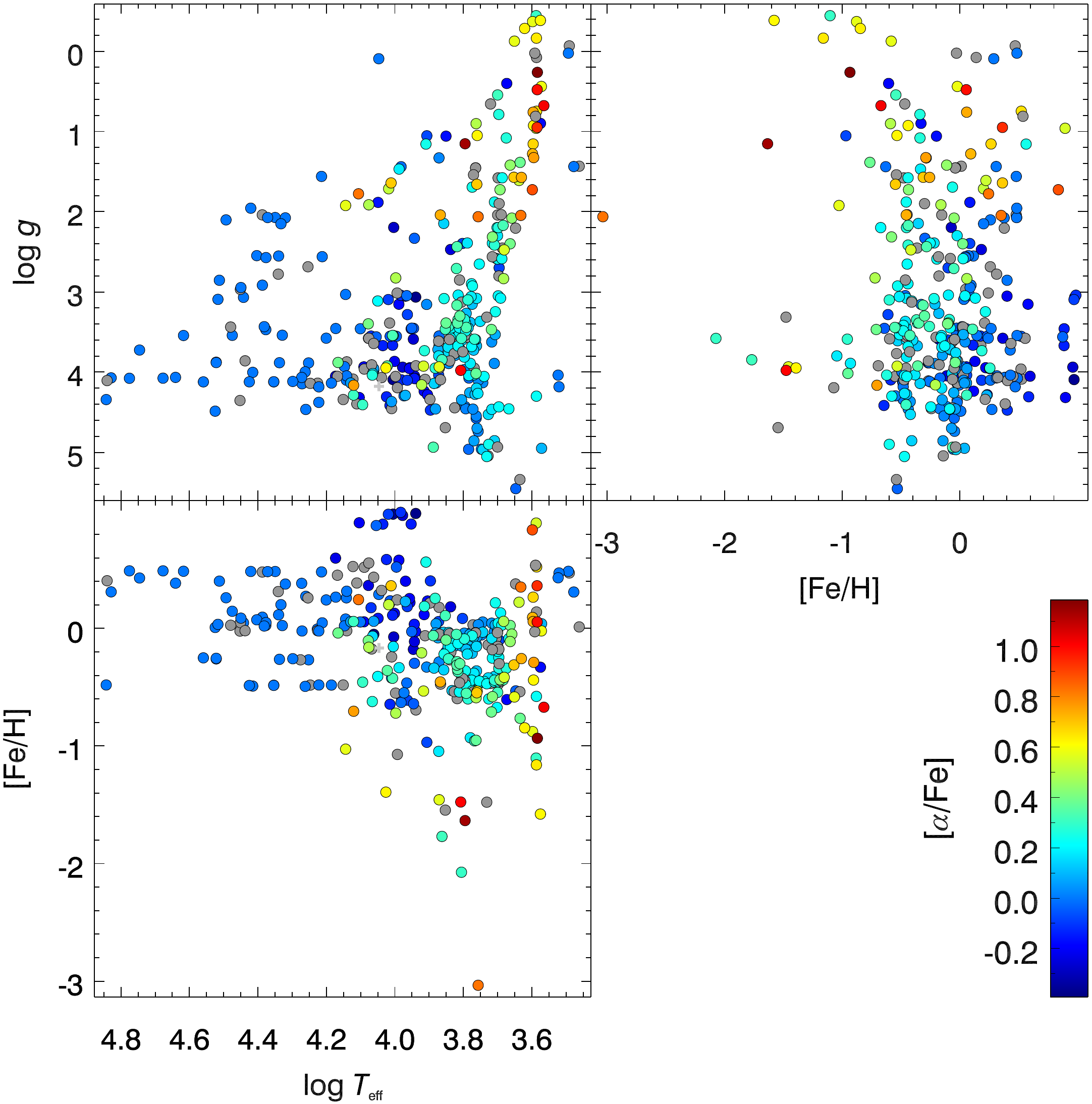}
\caption{$T_{\mathrm{eff}}$, [Fe/H], and $\mathrm{log}~g$ coverage of the UVES-POP stellar library with [$\alpha$/Fe] color-coded. The points in grey show the stars for which [$\alpha$/Fe] was not determined.}
\label{plot_param}
\end{figure}

\begin{figure}
\centering
\includegraphics[width=1\hsize]{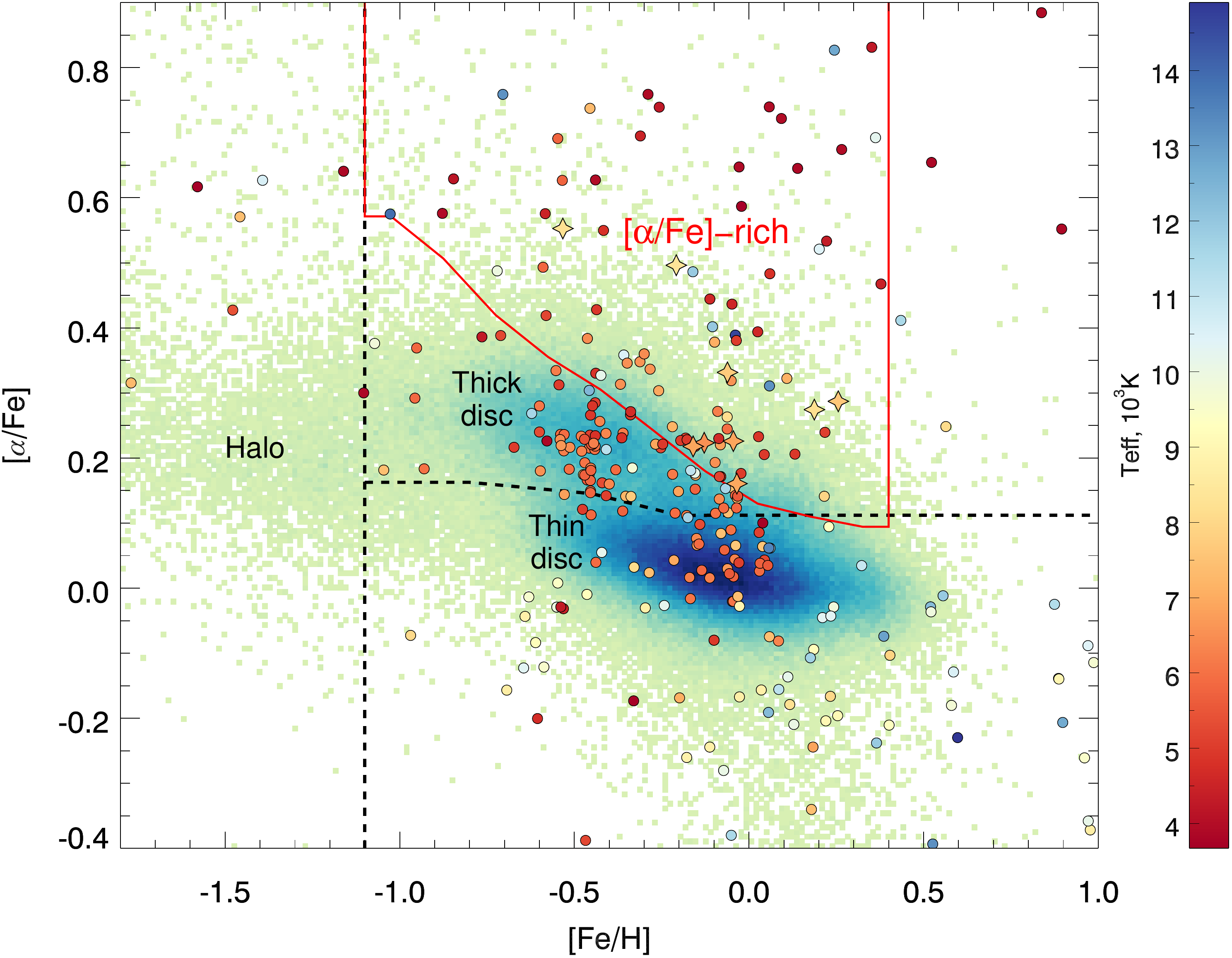}
\caption{Positions of the UVES-POP stars in the [$\alpha$/Fe]-[Fe/H] plane with $T_{\mathrm{eff}}$ color-coded. Only the stars with  [$\alpha$/Fe] determined in our study are shown. The underlying number density plot shows positions of stars from the 3rd release of the GALAH survey \citep{buder2021} with the high SNR and quality criteria for both [$\alpha$/Fe] and [Fe/H] applied to the input catalog. Black dashed line shows a separation between thin and thick discs obtained by considering the bimodal [$\alpha$/Fe] distribution at different [Fe/H] that will be described in detail in Borisov et~al. (in prep.) where the authors also assume that objects of the Galactic halo have [Fe/H]$\leq$-1.1~dex. The red solid line delineates the area that contains extremely [$\alpha$/Fe]-rich stars (see \S\ref{subsec:fit}), the candidate ``y-ex$\alpha$fe'' stars are shown as four-pointed stars.}
\label{plot_afe_vs_feh}
\end{figure}

To use stellar spectra from the recalibrated library for stellar population synthesis, it is crucial to determine the fundamental parameters of stellar atmospheres ($T_{\mathrm{eff}}, \log g$, [Fe/H], [$\alpha$/Fe], $v \sin i$). Literature data cover only about half of the stars of the UVES-POP library \citep{ambre_uves,ambre_harps,pastel,x-shooter-dr2_param,indo-us}. Moreover, the measurements are heterogeneous, i.e., determined with different algorithms having different systematic problems that are applied to spectra of varying quality originating from many observing facilities. Also, the flux recalibration, which we performed could change the values of parameters in some stars even if the literature measurements were made on UVES data. Therefore, we decided to determine them for all stars (where possible) using the same technique.

We decided to measure the fundamental stellar parameters fitting the full spectrum against a grid of synthetic stellar atmospheres because all existing grids of empirical stellar spectra have their own systematics related to (i) interpolation of stellar spectra irregularly placed in the parameter space onto a regular grid; (ii) propagation of potentially biased values of parameters from one dataset to another, e.g., when parameters of one stellar library (X-Shooter Spectral Library; \citealp{x-shooter-dr2_param}) are based on the calibration obtained from another one (ELODIE-3.1; \citealp{elodie_new_2}) using a previous version of the calibration of the same method. We also intentionally decided not to use a technique based on the interpolation of spectra in the parameter space to avoid additional uncertainties related to the choice of the interpolation method.

\subsection{A hybrid minimization algorithm}
\label{subsec:method}

High signal-to-noise ratios, good calibration quality of the reprocessed UVES-POP spectra, and the wide variety of spectral classes require the use of modern high-quality stellar models having as complete coverage of the parameter space as possible. We chose the PHOENIX synthetic models \citep{phoenix} computed over a wide range of values in the parameter space ($2300 < T_{\mathrm{eff}} < 15,000$~{K}, $-0.5 < \mathrm{log}~g < 6.5$, $-4 < \mathrm{[Fe/H]} < 1$~dex for 8 different values of [$\alpha$/Fe] from $-0.4$ to $+1.2$~dex) and complemented them with the models from the older BT-Settl library \citep{bt-settl} in the high- ($> 15,000$~K) and low temperature ($< 3,000$~K) regimes. For the BT-Settl models, we fixed [$\alpha$/Fe]=0~dex. This, however, does not contradict the standard theory of the Galactic evolution because hot (i.e. massive) stars are generally not $\alpha$-enriched \citep[e.g., ][]{Mishenina2004}. The combined grid of models has the full wavelength coverage ($300 < \lambda < 2500$~nm) at the spectral resolution $R = 20,000$, but with rather complex and irregular node positions in the 4-dimensional parameter space ($T_{\mathrm{eff}}$, $\mathrm{log}~g$, [Fe/H], [$\alpha$/Fe]).

The currently available minimization algorithms for finding a solution, which can be integrated into a spectrum fitting technique do not handle correctly such complex model grids without interpolation onto a regular grid, which is subject to numerical artifacts. The stellar radial velocity and a projected  velocity of rotation, as well as a multiplicative continuum represented by a low-degree polynomial that absorbs potential flux calibration differences between synthetic and observed spectra also have to be determined. 

Therefore, we decided to develop a new hybrid minimization method \citep{hybmin_method} and implemented it as a function in IDL/GDL (the {\sc python} implementation is underway). Here we provide a brief description of our approach, its detailed description will be published in a separate paper (Rubtsov et al. in prep). The algorithm is designed to work with multidimensional grids of models without interpolation in the process of finding the best-fitting solution. It comprises the two main parts: (i) a hill climbing part which finds the minimum value of $\chi^2$ at a grid node using the connectivity matrix between the points in the parameter space computed by the triangulation part of the {\sc qhull} algorithm \citep{qhull}; (ii) finding the off-node position of the local $\chi^2$ minimum by performing a local approximation of the $\chi^2$ profile by a positively-defined quadratic form \citep{math_quad}. In the second part, a set of nodes (a certain simplex and all nodes associated with it) is selected so that the desired solution is inside the selected simplex. However, if the solution turns out to be outside a given simplex, then it is replaced by an adjacent one in the direction of the obtained solution, and the approximation is repeated. This happens until either the solution is found or the simplices are repeated (i.e. the algorithm loops). In the latter case, the simplices are merged, increasing the region of the parameter space available for finding the local minimum. In the worst-case scenario of a complex shape of the $\chi^2$ profile, the parameters from the first step will be considered to be the solution (and the corresponding flag is returned by the algorithm). This approach allows us to deal with both regular and irregular grids of models, but it can drive solutions into isolated local minima as most local minimization techniques.

This algorithm works with a discrete set of models distributed in a multidimensional parameters space and does not require interpolation between them. The hybridity consists in using an additional minimizer (in our case {\sc MPFIT}, \citealp{mpfit}) to determine independent continuous parameters at the requested grid nodes at each step (thus, the algorithm belongs to the ``greedy'' class). It also assumes that the continuous parameters determined at every step of the hill climbing process are not strongly degenerated with those determined discretely. During the full spectrum fitting of a stellar spectrum, 6 parameters are simultaneously determined: $T_{\mathrm{eff}}$, $\mathrm{log}~g$, [Fe/H], [$\alpha$/Fe] (discrete); $v_{rad}$, $v \sin i$, and a multiplicative polynomial continuum (continuous). In this particular case, the discrete and continuous parameters are not pairwise correlated. To illustrate how the method works, Fig.~\ref{method_procyon} shows a simplified version of the fit using two free discrete parameters.

\subsection{Stellar atmospheric parameters}
\label{subsec:fit}

For all stars in the recalibrated library, we have collected the available information on atmospheric parameters, spectral types, and variability. The resulting dataset turned out to be very heterogeneous (45 individual data sources), but nonetheless, these data can be used for selection criteria and initial guess for the spectrum fitting procedure using our technique for the determination of the fundamental stellar parameters.

From the full set of stars in the UVES-POP library (406 stars), we excluded carbon stars (6 C), hot stars with emission lines and very few absorption lines (16 AeBe), and Wolf-Rayet stars (8 WR). Also, we excluded spectra consisting of only one atmospheric UV segment 346B (12 spectra). In total, 364 spectra remained in the sample for the determination of the stellar atmospheric parameters.

We fitted 244 of the 364 stars in the wavelength range $380-680$~nm, which is relatively well described by synthetic models and allows us to directly compare our results with those obtained in a similar fashion from other empirical libraries of stellar spectra in the same wavelength range (e.g., ELODIE, INDO-US, X-Shooter). For 120 low-temperature stars ($T_{\mathrm{eff}} < 4500$~K), high-temperature stars ($T_{\mathrm{eff}} > 12000$~K) and stars with missing spectral segments, the fit was carried out in the entire available wavelength range ($320-1025$~nm). Additionally, to improve the convergence and stability of the fitting and using assumptions from the stellar physics, we limited the rotational velocity to $v \sin i<15$~km s$^{-1}$ for giant stars with $\mathrm{log}~g < 2.75$. We also generated masks based on the initial guess of the spectral type, which excluded poorly modeled absorption lines in the PHOENIX models (diffuse interstellar bands and helium in hot stars, lithium in cool stars), cores of NaD and Ca H\&K lines potentially affected by interstellar absorptions (and chromospheric activity for Ca H\&K lines in cool dwarfs), cores of emission Balmer lines in AeBe stars, which were retained in the spectrum fitting sample and also in cool emission-line giants (Me).

Fig.~\ref{fit_examples} shows several examples of parameter determination using the full spectrum fitting technique described above for stars having different spectral types and luminosity classes. Figures~\ref{plot_param}--\ref{plot_afe_vs_feh} show the projections of the distribution of the obtained parameters $T_{\mathrm{eff}}$, $\mathrm{log}~g$, [Fe/H], [$\alpha$/Fe]. Among our sample stars, we found several dwarfs that belong to a rare type of object recently discovered and described in detail in \citet{Borisov2022} referred to as ``y-ex$\alpha$fe'' stars. These stars have abnormally high [$\alpha$/Fe] ratios, given their young ages. Following the criteria for the young age$\leq$3~Gyr and using the available age determination from the literature, we identified nine stars that are very likely ``y-ex$\alpha$fe'' candidates. Seven of them also have estimates of the eccentricity of their orbits in the Galaxy, which is originally considered as one of the young age indicators. The eccentricities of these stars are low and lie within the range for ``y-ex$\alpha$fe'' shown in fig.~6 in \citet{Borisov2022}. The red line in Fig.~\ref{plot_afe_vs_feh} delineates the locus of ``y-ex$\alpha$fe''s and the new candidates are shown as four-pointed stars. 

Using the stellar atmospheric parameters and luminosities, we estimated stellar masses and ages for 217 stars with the tool {\sc SPInS} \citep{Lebreton2020}. To compute luminosities, we applied the same procedure as in \citet{Borisov2022} except for the estimation of bolometric correction that was computed according to \citet{Creevey2022}.

\subsection{Comparison with other libraries}
\label{subsec:comp}

\begin{figure*}[]
\centering
\includegraphics[width=1\hsize]{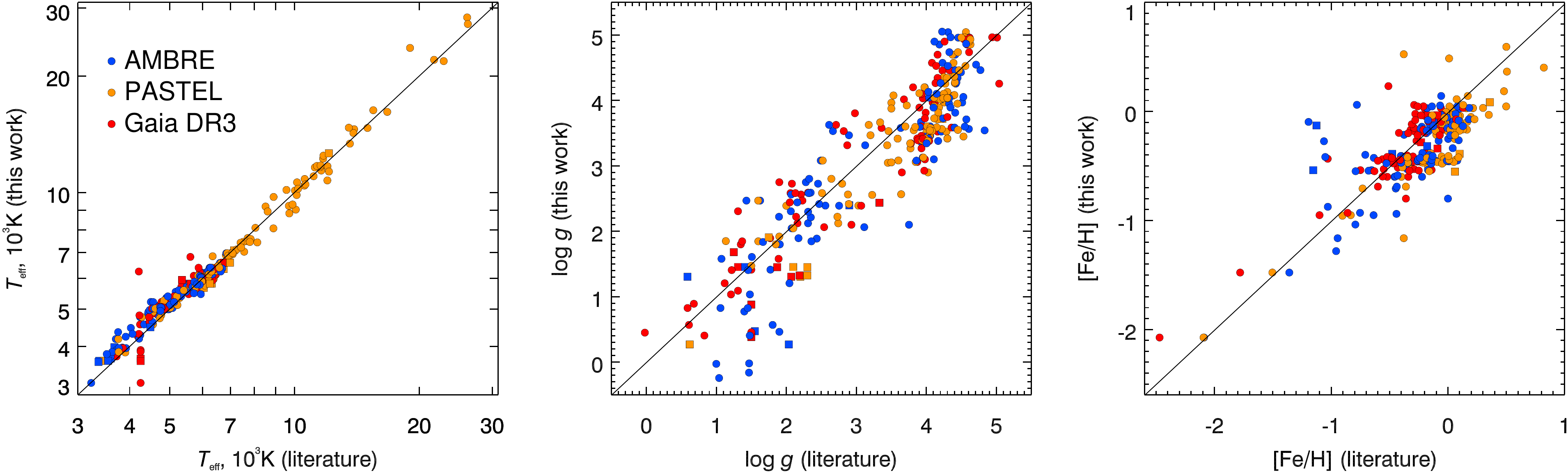}
\caption{Comparison of atmospheric parameters $T_{\mathrm{eff}}$, log~$g$, and [Fe/H] between this work and the literature. Stars that are presented in more than one catalog are shown multiple times. Pulsating variable stars are marked by squares.}
\label{uvs_param_comp_plot}
\end{figure*}

The largest samples of literature measurement on fundamental stellar parameters for the stars matching our recalibrated UVES-POP list are found in the AMBRE \citep{ambre_uves,ambre_harps}, PASTEL \citep{pastel}, and \textit{Gaia}~DR3 \citep{gaia_dr3,Recio-Blanco2022} catalogs, 123, 174, and 91 stars respectively (only objects with available measurements counted). We compare our parameters with these datasets in Fig.~\ref{uvs_param_comp_plot}. The AMBRE catalog includes only FGKM stars, whereas PASTEL also contains early-type A and B stars. When we compare with the \textit{Gaia}~DR3 data, we exclude stars with $T_{\mathrm{eff}}^\mathrm{Gaia}>7000$~K since their $T_\mathrm{eff}$ values show systematic underestimation. One should keep in mind, that AMBRE measurements were obtained homogeneously by applying the same data analysis technique to UVES, HARPS and FEROS archival data, while PASTEL is a literature compilation. Our $T_\mathrm{eff}$ measurements show good agreement with the literature data: the relative RMS$_{T\mathrm{eff}}$ values are around 3.9\% when comparing with AMBRE, $\sim$4.2\% for PASTEL, and $\sim$4.8\% for \textit{Gaia}~DR3. Comparison of log~$g$ gives the following values: RMS$_{\mathrm{log}~g}^\mathrm{AMBRE}$=0.62~dex, RMS$_{\mathrm{log}~g}^\mathrm{PASTEL}$=0.32~dex, and RMS$_{\mathrm{log}~g}^\mathrm{Gaia}$=0.51~dex. Finally, we compared metallicities of stars: RMS$_\mathrm{[Fe/H]}^\mathrm{AMBRE}$=0.30~dex, however, such a high value is caused by large discrepancy for giants/supergiants with low values of log~$g$. If we consider only stars with log~$g\geq$3, the agreement improves drastically:  RMS$_\mathrm{[Fe/H]}^\mathrm{AMBRE}$=0.16~dex. For PASTEL,  RMS$_\mathrm{[Fe/H]}^\mathrm{PASTEL}$=0.20~dex, however there is a systematic shift of $\sim-0.17$~dex. Comparison with \textit{Gaia} gives approximately the same level of agreement: RMS$_\mathrm{[Fe/H]}^\mathrm{Gaia}$=0.19~dex.

An important factor that might affect the comparison is variability of stars. We have already provided some statistics on them in subsection~\ref{subsec:var_stars}. In Fig.~\ref{uvs_param_comp_plot} we show pulsating variables as squares and note that they might change their parameters significantly due to variations of $T_\mathrm{eff}$ and log~$g$.

We have also compared calibrated flux densities in our spectra with those from the spectral libraries NGSL \citep{ngsl}, X-Shooter~DR3 \citep{x-shooter-dr3}, and ELODIE~3.1 \citep{elodie_new}. We cross-matched our list of stars with these three libraries and found 12, 8, and 25 stars in common with each of them correspondingly. For comparison purposes, we matched the spectral resolution of the spectra being compared by convolving one of them with a Gaussian with the width corresponding to the quadratic difference of the spectral resolution values. Figures~\ref{comp_ngsl}-\ref{comp_el} in the Appendix demonstrate the results of the comparison.

We emphasize the comparison with NGSL which contains spectra collected with the Space Telescope Imaging Spectrograph (STIS) operated at the Hubble Space Telescope. This library has excellent quality of flux calibration because of the location of the telescope outside the Earth's atmosphere. Fig.~\ref{comp_ngsl} demonstrates excellent agreement of 1.0--3.5\%\ in 10 stars out of 12. The worst agreement is in HD~102212, which is a Mira-type long periodic pulsating variable, therefore the discrepancy can occur from the observations being collected at different pulsation phases. Surprisingly, another long periodic variable, HD~206778 shows very good agreement between the two spectral libraries. For HD~47839 (and much less obvious for HD~63077 and HD~76932) we see a disagreement in the atmospheric UV part bluer of the Balmer break: we suspect that the origin of this problem is in the imperfect sensitivity curve determination and flux calibration, which uses tabulated atmospheric extinction coefficients for Paranal, which might change significantly at short wavelengths. We also note, that 2 stars, HD~111786 and HD~142703 are $\delta$~Sct type pulsating variables, which might explain the spectral differences in the region containing many high-order $\log g$-sensitive Balmer lines. There is another variable star, HD~22049 of the \emph{BY~Dra} type (i.e. spotted late-type dwarf) which does not exhibit substantial spectral differences between UVES-POP and NGSL. In Table~\ref{ngsl_rms}, we present the mean relative RMS for each spectral segment of UVES spectra compared to NGSL from all 12 stars (2nd column) and from 11 stars excluding HD~102212 (3rd column).

\begin{deluxetable}{lcc}
\tablecaption{Mean values of the relative RMS of flux difference between UVES-POP and NGSL for each spectral segment.\label{ngsl_rms}}
\tablewidth{0pt} 
\tablehead{
\colhead{Dichroic} & \colhead{$<$RMS$_{12}>$} & \colhead{$<$RMS$_{11}>$}
}
\startdata
346B	& 4.3\% & 3.5\%  \\
437B	& 2.0\% & 2.0\%  \\
580L	& 1.8\% & 1.8\%  \\
580U	& 1.1\% & 1.0\%  \\
860L	& 1.1\% & 0.8\%  \\
860U	& 2.4\% & 1.8\%  \\
\enddata
\tablecomments{The middle column $<$RMS$_{12}>$ presents the values computed for all 12 matching spectra; the long periodic variable HD~102212 was excluded while computing the values presented in the right column $<$RMS$_{11}>$.}
\end{deluxetable}

In the case of the XSL~DR3 library, we compared our spectra with the second spectral segment of their spectra (the so-called VIS arm, 5400-10200~\AA, see \citet{x-shooter-dr3} for details). This comparison shows a somewhat worse agreement (Fig.~\ref{comp_xsl}), especially in the blue end of the spectral range affected by the `wavy' throughput of the dichroic mirror, which, however, is masked in the merged X-Shooter spectra presented in the public release. 

Comparison with ELODIE~3.1 (Fig.~\ref{comp_el}) shows the worst agreement among the three aforementioned libraries. We can see relatively good correspondence in the center of the range, however, in many cases, there is a strong ``bending'' of flux at the edges, especially in the blue part.

Neither X-Shooter nor ELODIE used broad- and/or middle-band photometry to correct the global shape of the spectra. We, therefore, attribute the disagreement to these stellar libraries rather than to our collection of spectra.

\begin{figure*}[]
\centering
\includegraphics[width=0.49\hsize]{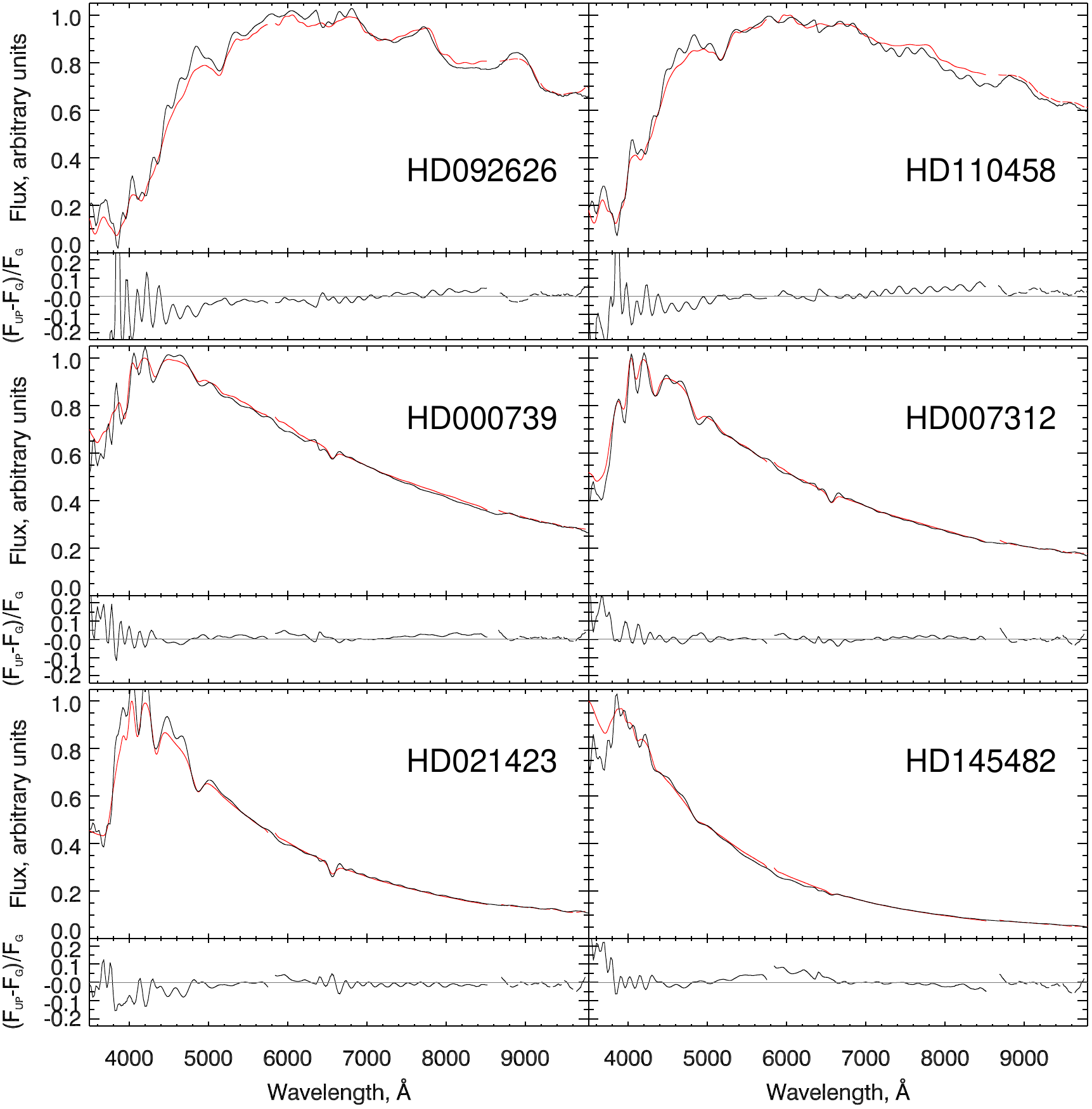}
\includegraphics[width=0.49\hsize]{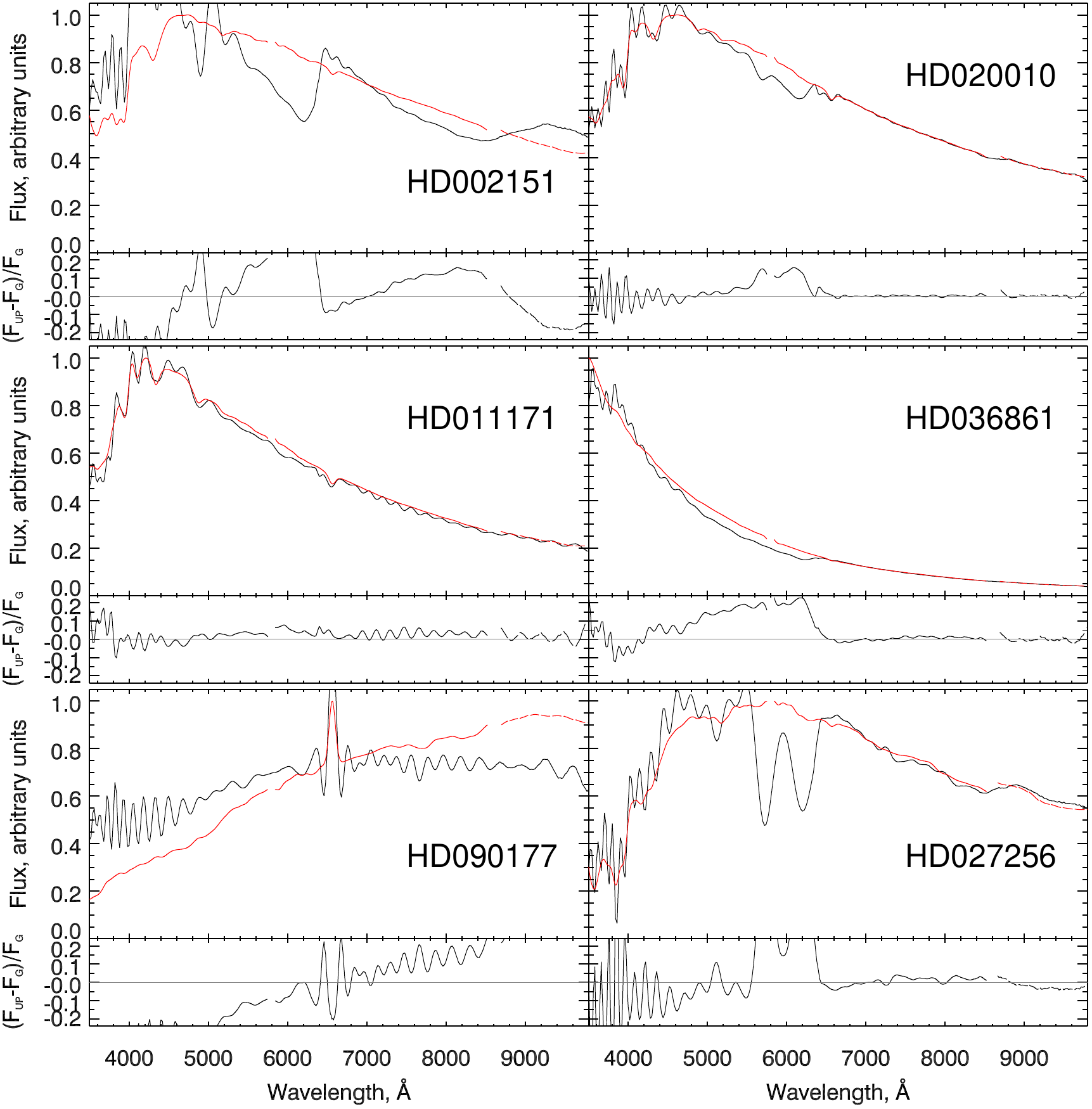}
\caption{Examples of 12 resolution-degraded UVES-POP spectra (red) with \textit{Gaia} BP/RP spectra (black) overplotted. All spectra are normalized to unity at the maximal  flux of the convolved UVES-POP spectrum. Below each plot we show a panel with the fractional flux differences between the two spectra. The left 6 panels present the stars with a satisfactory agreement between UVES-POP and \textit{Gaia}; the right 6 panels show stars with obvious calibration issues in the \textit{Gaia} BP/RP spectra. None of the 12 stars has significant variability according to SIMBAD.}
\label{gaia_bprp_spec_fig}
\end{figure*}

While we were finalizing the work on this paper, the \textit{Gaia} collaboration released low-resolution BP/RP spectra as a part of DR3 \citep{gaia_dr3,gaia_spectra_bp_rp}. 302 stars from our sample have \textit{Gaia} BP/RP spectra, which we directly compared with our data. We first convolve UVES spectra with the line spread function of \textit{Gaia} BP/RP spectra \citep[see fig.~17 in][]{gaia_spectra_bp_rp} varying across the wavelength. One should keep in mind that because of the extreme difference in spectral resolution the convolution kernel is very broad and it might cause edge effects in the comparison. Then we bin the convolved spectra in the spectral segments, which correspond to spectral pixels in \textit{Gaia} data. Finally, we directly compare the spectral flux densities in each pixel. In Fig.~\ref{gaia_bprp_spec_fig} we compare resolution-degraded UVES spectra with BP/RP spectra from \textit{Gaia}~DR3 for several stars. While for many spectra the overall continuum shape matches between two sources (Fig.~\ref{gaia_bprp_spec_fig}, left 6 panels), about 1/3 of the BP/RP spectra display strong continuum artifacts, the most frequent being moderate to strong `dips' in the red end of the BP \textit{Gaia} segment at $\lambda$=500--670~nm (Fig.~\ref{gaia_bprp_spec_fig}, right 6 panels). This likely indicates an issue in the spectrophotometric calibration of \textit{Gaia} low-resolution spectra because none of the other sources, which we compared UVES spectra with, does not display similar features. Besides, in the majority of \textit{Gaia} BP/RP spectra even when the overall continuum shape matches that of UVES spectra, we see high-frequency `waves' (peak-to-peak distance of $\sim$20~nm) with amplitudes reaching 20~\%. Therefore, we conclude that at present one should not use \textit{Gaia}~DR3 BP/RP spectra as tertiary spectrophotometric standards to perform flux calibration of higher resolution spectra. Hopefully, these spectrophotometric calibration issues will be corrected in the next \textit{Gaia} Data Release.

\section{Web-service for data access and quality control} 
\label{sec:web}

\begin{figure*}[]
\centering
\includegraphics[width=\hsize]{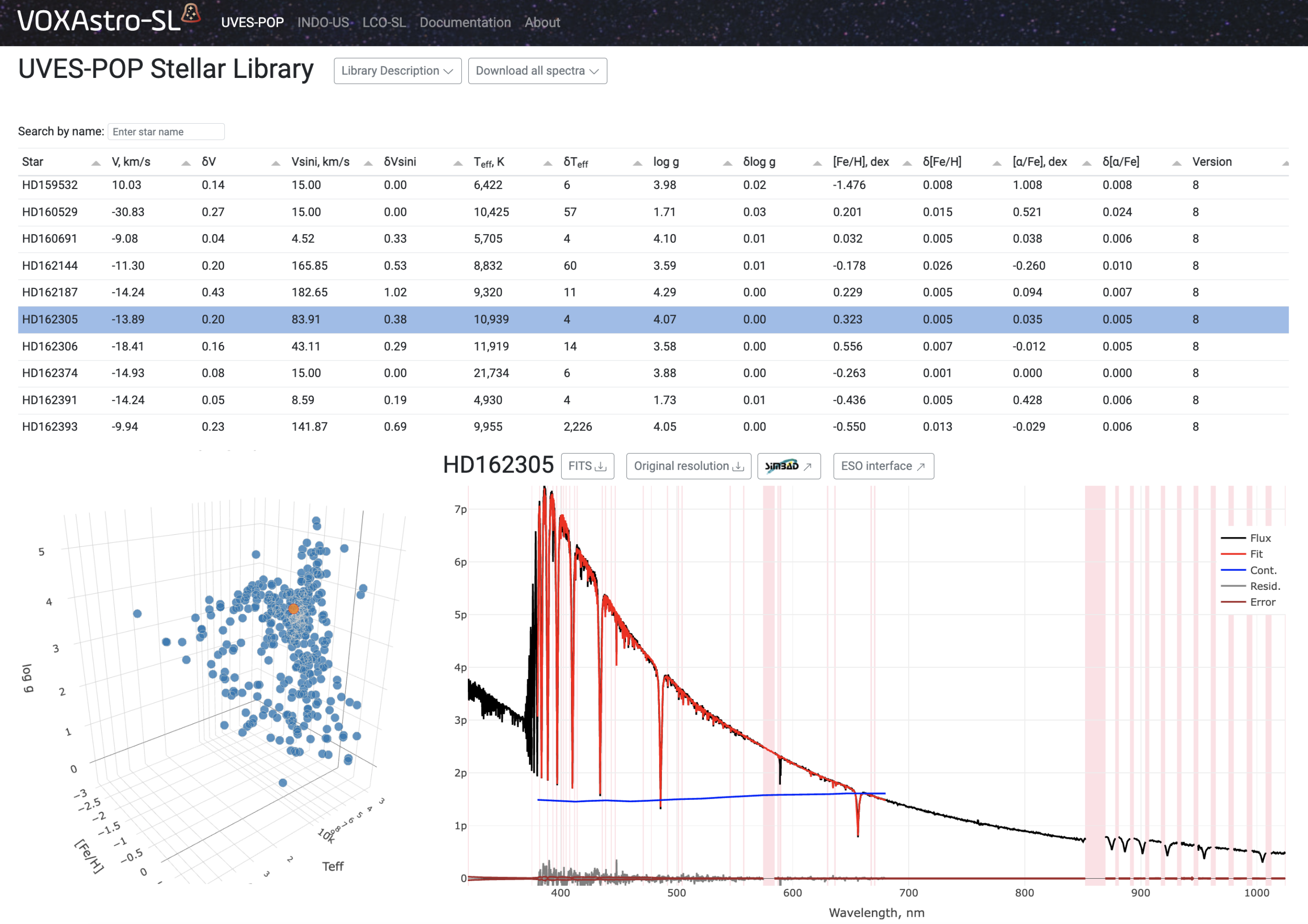}
\caption{
A screenshot of an interactive web-based spectral viewer.
An R=20,000 spectrum of HD~162305 is shown in black, the best-fitting model is in red, the multiplicative polynomial continuum is in blue. Pink-shaded areas denote the masked regions excluded from the fit.}
\label{fig_viewer}
\end{figure*}

We publish the results of the recalibration and analysis of the spectra in the UVES-POP library in the dedicated web-site of the ``VOXAstro Stellar Libraries'' project: \url{https://sl.voxastro.org/library/UVES-POP/details/}. We created a user-friendly modern and interactive web-interface for accessing and visualizing data.
Server-side part of the application uses Python-based framework Django\footnote{\url{https://www.djangoproject.com/}} on top of the PostgreSQL database.
Interactive frontend part of the application was developed using JavaScript framework Vue.js\footnote{\url{https://vuejs.org/}} and Plotly.js library\footnote{\url{https://plotly.com/javascript/}} which allows us to efficiently display high-resolution spectra. A screenshot of this tool with a spectrum of HD~162889 is shown in Fig.~\ref{fig_viewer}.

In the top section of the web page, there are links to the general description of the library, buttons for downloading all spectra at once in a single archive (see the description of the FITS binary table structure of the output results in table~\ref{tab_fitspec}) and/or the table with stellar atmospheric parameters (see table~\ref{tab_fitres}) in the form of an interactive table. It allows one to quickly find an object of interest in the database and sort the results by different parameters. This table is related to the 3D visualization of the parameter space for stellar atmospheres in the bottom left part of the page. This interactive 3D plot in the $T_{\mathrm{eff}}$, $\mathrm{log}~g$, [Fe/H] space displays the results of the full spectrum fitting of UVES-POP spectra performed by our data analysis technique described above.

Selection of a data point in this plot or a line in the data table above will display a spectrum with its best-fitting results on the bottom right panel of the web-page as well as a download button for that particular spectrum and a link to the SIMBAD database and ESO archive to the data for the corresponding star.

We developed a dedicated password-protected part of the project web-site to perform (i) quality control for the data reduction, order and segment merging, telluric correction, flux calibration; (ii) assessment of the full spectrum fitting quality using synthetic stellar atmospheres; (iii) reliability of the derived stellar atmospheric parameters. Each of the three aspects is graded from 1 (inadequate) to 5 (excellent). Users performing quality control can also leave comments and see a full version history of the recalibration for every star. Most stars in the current release are at version~7. After each iteration of the quality control corresponding to a separate version number, the spectra, which were found to have problems with at least one of the 3 aspects, were re-processed. To minimize the subjective factor of the grading, several members of our team were involved in the quality control. At least three persons checked each spectrum (and a majority of spectra were checked by 5 individuals). 

To grade the recalibration quality, a reviewing person had to: (i) inspect a spectrum for the presence of residual ripples between Echelle orders and flux ``steps'' between UVES spectral segments; (ii) inspect the regions of strong telluric absorption for strong residuals; (iii) assess the global spectral shape to detect large-scale flux calibration errors. To grade the full spectrum fitting quality, one had to: (i) inspect fitting residuals in the wings of Balmer lines in intermediate-to-hot stars; (ii) verify whether Balmer line cores were properly modeled (does not apply to emission-line stars); (iii) check whether molecular bands in cool stars were adequately described by the model; (iv) check the residuals in several known multiplets of iron and the most prominent absorption lines of $\alpha$-elements (e.g Mg$b$) to assess the quality of the derived iron- and $\alpha$-abundances. Finally, to grade the quality of the stellar atmospheric parameters one had to check the availability of literature data in the SIMBAD database or at least a spectral type and a luminosity class and conclude whether the derived parameters corresponded to the published spectral type and/or atmospheric parameters. For example, if a star is classified as A5V in the literature, and its derived $\log g = 3.0$, it likely indicates a problem.

From the results of the quality control procedure, we conclude that (i) the spectrum recalibration quality is generally very good except 2--3 stars where an unknown factor caused the distortion of flux in one of the blue segments; (ii) the full spectrum fitting quality against a grid of synthetic stellar spectra is the best at sub-solar metallicities in the range of $4800<T_{\mathrm{eff}}<13000$~K, it deteriorates at lower temperatures and higher metallicities potentially because of incomplete spectral line lists and/or wrong opacities used in the stellar atmospheric modeling for cool stars and also at high temperatures, where non-LTE effects become important; (iii) stellar atmospheric parameters are sometimes biased because of degeneracies, i.e. slightly overestimated $T_{\mathrm{eff}}$ leads to significantly overestimated $\log g$, however, the use of the novel minimization technique has drastically reduced these effects.

\section{Fitting galaxy spectra using UVES-POP stellar templates}
\label{sec:stpop}

\begin{figure}
\centering
\includegraphics[width=1\linewidth]{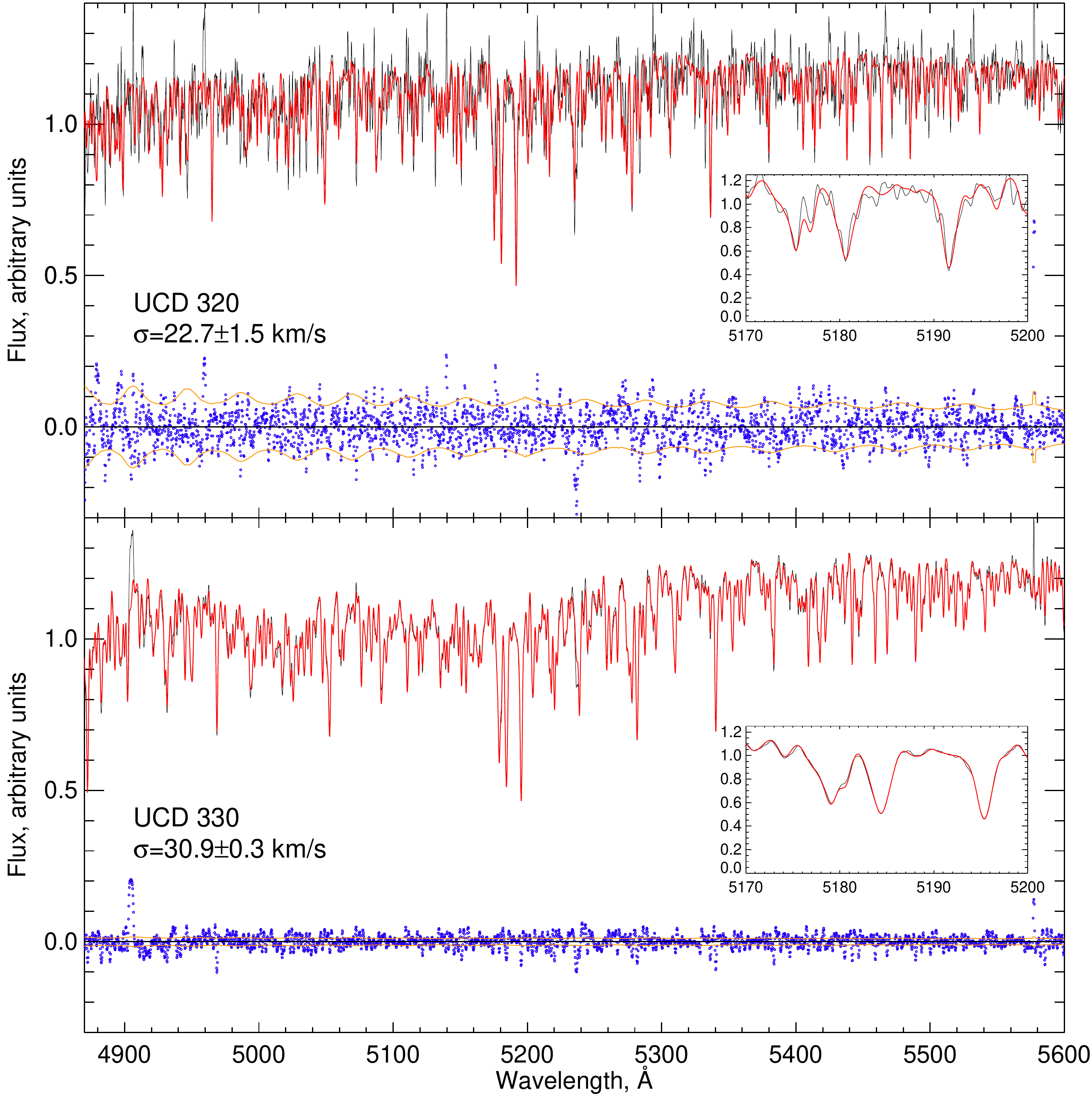}
\caption{Spectra of UCD~320 and UCD~330 (black) and best-fitting models (red), residuals are shown as blue points. The insets show the spectral region around the Mg$b$ triplet. For better visibility, flux errors and residuals for UCD~320 are reduced by a factor of 4.}
\label{ucd_plot}
\end{figure}

As an additional test of the quality of the UVES-POP spectral library recalibration, we performed a pixel space fitting of the two ultra-compact dwarf galaxies UCD~320 and UCD~330 in the nearby Centaurus~A galaxy group in order to estimate their radial velocities and stellar velocity dispersions. Such galaxies typically host old stellar populations \citep{2008MNRAS.390..906C,2011MNRAS.412.1627C} often enhanced in $\alpha$-elements \citep{2012MNRAS.425..325F}, in which the prevailing contribution to the total luminosity in the optical domain comes from G- and K-giants. Although UCDs are usually described by simple stellar populations, they often show color gradients in high-resolution images, which might suggest a radial metallicity gradient \citep{a_afanasiev}. Therefore, it is reasonable to use several template stars spanning a range of metallicities to infer stellar kinematics of UCDs.

We selected 8 recalibrated spectra of G and K giants with the atmospheric parameters lying in the range $4000 \leq T_{\mathrm{eff}} \leq 6000$~K, 1 $\leq$ log~$g \leq$ 2 and -1 $\leq$ [Fe/H] $\leq$ 0~dex. Spectra of both UCDs obtained with UVES were downloaded from the ESO Archive. These spectra were taken with 1-arcsec slit and have the resolving power of about R$\approx$42,000. We degraded the spectral resolving power in the UCD spectra to R=20,000 to match the recalibrated UVES-POP stellar spectra. This also improved signal-to-noise ratios in these spectra: the final SNR values after the convolution per pixel are SNR$_{320}\!\!\approx$4 and SNR$_{330}\!\!\approx$24. To estimate radial velocity and velocity dispersion, we used the penalised pixel fitting procedure (pPXF, \citealp{ppxf}) with a set of 8 templates. The values we obtained are $\sigma=22.7\pm1.5$~km s$^{-1}$ and $\sigma=30.9\pm0.3$~km s$^{-1}$ for UCD~320 and UCD~330 respectively. They are in a perfect agreement with \citet{k_voggel}, \citet{m_rejkuba}, and \citet{m_taylor} (UCD~320) and \citet{k_voggel}, \citet{m_rejkuba}, and \citet{s_hernandez} (UCD~330). The results of the spectral fitting for the two UCDs are shown in Fig.~\ref{ucd_plot}.

\section{Summary}

We presented the results of re-processing, recalibrating and analysing a data set containing 406 high-resolution stellar spectra with R=20,000 and R=80,000 from the UVES-POP stellar spectral library. The main goal of this project is to obtain a set of well-calibrated spectra, which can be used for stellar population synthesis. Our results are as follows:
\begin{itemize}
\item We solved the problem of imperfect Echelle order matching. The ``distortion'' of Echelle orders in flux space was producing ripples in the regions of the merged spectra where the orders overlap. This prevented the spectra from the original data release to be used for stellar population synthesis.
\item We performed the merging of non-overlapping spectral segments by using synthetic stellar atmospheres. The original UVES-POP spectra equated the fluxes from each side of the wavelength gaps leading to ``steps'' in the overall spectral shape.
\item We performed a telluric correction for all spectra using a dedicated algorithm using a grid of models computed using the {\sc ESO~SkyCalc} tool. This eliminates telluric absorption features produced by oxygen, water vapor and ozone in Earth's atmosphere. 
\item We performed the spectro-photometric calibration of UVES-POP spectra by using 9 photometric catalogs including two catalogs from the Tycho and \textit{Gaia} space missions. The final spectra are converted into absolute physical flux density units erg~cm$^{-2}$~s$^{-1}$~\AA$^{-1}$, fluxes are corrected to above the Earth's atmosphere.
\item We estimated interstellar extinction values for 41 stars using 3D dust extinction maps and for 364 stars $E(B-V)$ values were calculated from the fitting of atmospheric parameters.
\item We matched our sample with the General Catalog of Variable Stars and discussed potential problems of spectral recalibration which may arise from intrinsic variability.
\item We compared the recalibrated UVES-POP spectra for a sub-sample of stars with the spectra from three other spectral libraries (NGSL, ELODIE, X-Shooter) to conclude that our photometry-based flux calibration approach was successful.
\item For 364 spectra, we computed fundamental stellar atmospheric parameters $T_{\mathrm{eff}}$, $\mathrm{log}~g$, [Fe/H], [$\alpha$/Fe] as well as radial velocity and rotational velocity ($v_{rad}$, $v \sin i$), using our own full-spectrum fitting technique based on PHOENIX/BT-Settl synthetic stellar atmospheres. The comparison of the derived parameters with published data from the AMBRE and PASTEL catalogs shows good agreement.
\item We presented an example of the usage of the recalibrated UVES-POP spectra by performing full-spectrum fitting of archival UVES spectra of two dwarf galaxies and estimated their stellar velocity dispersion, which happened to agree perfectly with the published values for these two galaxies.
\item All the data and spectrum fitting results are made publicly available from the project website \mbox{\url{https://sl.voxastro.org/library/UVES-POP/details/}}, which also provides convenient data visualization tools.
\end{itemize}

\smallskip
\noindent The initial part of this work and spectral recalibration was supported by the RScF grant 17-72-20119; the efforts on the spectrophotometric recalibration were supported by the RScF grant 19-12-00281. IC's research is supported by the SAO Telescope Data Center. SB, IC, and KG also acknowledge support by the ESO Visiting Scientist Programme. We thank the anonymous referee for the important comments that improved the quality of this paper. We are grateful to Albert Shaykhutdinov (ASC~LPI, Russia) for his help with the calculation of topocentric corrections and Patrick de~Laverny (OCA, France) for useful comments. This research made use of the SIMBAD database and VizieR catalog access tool, both operated at CDS, Strasbourg, France. 
\facilities{ESO VLT}

\bibliographystyle{aasjournal}
\bibliography{uves-pop.bib}

\appendix
\section{Description of the columns in the data tables.}

\begin{center}
\begin{longtable}{llll}
\caption{{Description of the output structure of the spectrum fitting results stored in the binary FITS table format}}
\label{tab_fitspec}
\\ \hline \hline
Column Name & Units & Datatype & Description \\
\hline \hline
\endfirsthead
\caption{{Description of the output structure of the spectrum fitting results (cont.)}}
\\ \hline \hline
Column Name & Units & Datatype & Description \\
\hline \hline
\endhead
\hline \multicolumn{4}{r}{{Continued on next page}} \\
\endfoot
\hline
\endlastfoot
objname &   & string &  UVES-POP identifier \\
wave & nm & double &  Array of wavelengths \\
flux & erg~cm$^{-2}$s$^{-1}$\AA$^{-1}$ & double &  Array of flux \\
error & erg~cm$^{-2}$s$^{-1}$\AA$^{-1}$ & double &  Array of flux error \\
pixmask &   & long &  Array of mask \\
quality &   & long &  Array of quality flags \\
swlvec & nm & double &  Array of wavelength scale shifts \\
lsfvec & km~s$^{-1}$ & double &  Array of LSF velocity and dispersion \\
model & erg~cm$^{-2}$s$^{-1}$\AA$^{-1}$ & double &  Stellar spectrum model from grid \\
mcont &   & double &  Multiplicative continuum \\
bestfit & erg~cm$^{-2}$s$^{-1}$\AA$^{-1}$ & double &  Best fit model \\
method &   & string &  Name of data analysis method \\
grid\_name &   & string &  Name of the model grid used \\
mdegree &   & long &  Degree of multiplicative polynomial \\
wlr & nm & double &  Wavelength range for fitting \\
i\_v & km~s$^{-1}$ & double &  Initial guess for radial velocity \\
v & km~s$^{-1}$ & double &  Radial velocity \\
e\_v & km~s$^{-1}$ & double &  Uncertainty of radial velocity \\
i\_vsini & km~s$^{-1}$ & double &  Initial guess for projection of the rotation velocity \\
vsini & km~s$^{-1}$ & double &  Projection of the rotation velocity \\
e\_vsini & km~s$^{-1}$ & double &  Uncertainty of projection of the rotation velocity \\
i\_teff & K & double &  Initial guess for effective temperature \\
teff & K & double &  Effective temperature \\
e\_teff & K & double &  Uncertainty of effective temperature \\
i\_logg & dex & double &  Initial guess for surface gravity \\
logg & dex & double &  Surface gravity \\
e\_logg & dex & double &  Uncertainty of surface gravity \\
i\_fe\_h & dex & double &  Initial guess for metallicity \\
fe\_h & dex & double &  Metallicity [Fe/H] \\
e\_fe\_h & dex & double &  Uncertainty of metallicity \\
i\_a\_fe & dex & double &  Initial guess for alpha-elements abundance \\
a\_fe & dex & double &  Alpha-elements abundance [$\alpha$/Fe] \\
e\_a\_fe & dex & double &  Uncertainty of alpha-elements abundance \\
chisqr &   & double &  $\chi^2$ statistics \\
dof &   & long &  Number of degrees of freedom \\
chi2dof &   & double &  Normalized $\chi^2$ \\
mpfit\_bestnorm &   & double &  Value of the summed squared weighted residuals \\
mpfit\_dof &   & long &  Computed number of degree of freedom \\
mpfit\_nfev &   & long &  Total number of function evaluations performed \\
mpfit\_nfree &   & long &  Number of free parameters in the fit \\
mpfit\_npegged &   & long &  Number of free parameters which are pegged at a limit \\
mpfit\_niter &   & long &  Total number of iterations completed \\
mpfit\_status &   & long &  Integer status code is returned (see MPFIT description) \\
mpfit\_errmsg &   & string &  String error or warning message is returned (see MPFIT description) \\
simbad &   & string &  SIMBAD identifier \\
hd &   & string &  HD identifier \\
hr &   & string &  HR identifier \\
hip &   & string &  Hipparcos identifier \\
tyc &   & string &  Tycho identifier \\
gaia\_dr2 &   & string &  \textit{Gaia}~DR2 identifier \\
gaia\_dr3 &   & string &  \textit{Gaia}~DR3 identifier \\
ra & deg & double &  Right Ascension \\
dec & deg & double &  Declination \\
date &   & string &  Date of observation (YYYY-MM-DDThh:mm:ss.sss) \\
exptime & s & double &  Total integration time \\
spclass &   & string &  Spectral class \\
mass & M$_{\odot}$ & double &  Stellar mass \\
e\_mass & M$_{\odot}$ & double &  Uncertainty of stellar mass \\
age & Gyr & double &  Stellar age \\
e\_age & Gyr & double &  Uncertainty of stellar age \\
ebv & mag & double &  Color excess E(B-V) from fit \\
ebv\_err & mag & double &  Uncertainty of E(B-V) from fit \\
ebv\_map & mag & double &  Color excess E(B-V) from DUSTMAPS \citep{2018JOSS....3..695G} \\
lit\_ref &   & string &  Reference to the literary source of measurements \\
lit\_v & km~s$^{-1}$ & double &  Radial velocity from literature \\
lit\_e\_v & km~s$^{-1}$ & double &  Uncertainty of radial velocity from literature \\
lit\_vsini & km~s$^{-1}$ & double &  Projection of the rotation velocity from literature \\
lit\_e\_vsini & km~s$^{-1}$ & double &  Uncertainty of projection of the rotation velocity from literature \\
lit\_teff & K & double &  Effective temperature from literature \\
lit\_e\_teff & K & double &  Uncertainty of effective temperature from literature \\
lit\_logg & dex & double &  Surface gravity from literature \\
lit\_e\_logg & dex & double &  Uncertainty of surface gravity from literature \\
lit\_fe\_h & dex & double &  Metallicity [Fe/H] from literature \\
lit\_e\_fe\_h & dex & double &  Uncertainty of metallicity from literature \\
lit\_a\_fe & dex & double &  Alpha-elements abundance [$\alpha$/Fe] from literature \\
lit\_e\_a\_fe & dex & double &  Uncertainty of alpha-elements abundance from literature \\
variance &   & long &  Variance flag (true/false) \\
gcvs &   & string &  GCVS identifier \\
gcvs\_n &   & string &  GCVS indicator \\
gcvs\_type &   & string &  Type of variance from GCVS \\
gcvs\_period & day & double &  Period of variance from GCVS \\
gcvs\_name &   & string &  Name of variance from GCVS \\
segments &   & long &  Segments flags \\
resolution &   & double &  Spectral resolution \\
phot\_name &   & string &  Array of filter names \\
phot\_mag & mag & double &  Array of magnitudes in corresponding filters \\
phot\_ref &   & string &  References to filters curves \\
\end{longtable}
\end{center}

\begin{center}
\begin{longtable}{llll}
\caption{Description of the structure with the catalog of atmospheric parameters}
\label{tab_fitres}
\\ \hline \hline
Column Name & Units & Datatype & Description \\
\hline \hline
\endfirsthead
\caption{Description of the structure with the catalog of atmospheric parameters (cont.)}
\\ \hline \hline
Column Name & Units & Datatype & Description \\
\hline \hline
\endhead
\hline \multicolumn{4}{r}{{Continued on next page}} \\
\endfoot
\hline
\endlastfoot
objname &   & string &  UVES-POP identifier \\
simbad &   & string &  SIMBAD identifier \\
hd &   & string &  HD identifier \\
hr &   & string &  HR identifier \\
hip &   & string &  Hipparcos identifier \\
tyc &   & string &  Tycho identifier \\
gaia\_dr2 &   & string &  \textit{Gaia}~DR2 identifier \\
gaia\_dr3 &   & string &  \textit{Gaia}~DR3 identifier \\
url\_r20 &   & string &  Link to download the spectrum R=20000 with analysis results \\
url\_r80 &   & string &  Link to download the spectrum R=80000 \\
ra & deg & float &  Right Ascension \\
dec & deg & float &  Declination \\
date &   & string &  Date of observation (YYYY-MM-DDThh:mm:ss.sss) \\
exptime & s & float &  Total integration time\\
spclass &   & string &  Spectral class \\
mass & M$_{\odot}$ & float &  Stellar mass \\
e\_mass & M$_{\odot}$ & float &  Uncertainty of stellar mass \\
age & Gyr & float &  Stellar age \\
e\_age & Gyr & float &  Uncertainty of stellar age \\
ebv & mag & float &  Color excess E(B-V) from fit \\
ebv\_err & mag & float &  Uncertainty of E(B-V) from fit \\
ebv\_map & mag & float &  Color excess E(B-V) from DUSTMAPS \citep{2018JOSS....3..695G} \\
nsegments &   & long &  Number of segments of spectrum \\
variance &   & long &  Variance flag (true/false) \\
gcvs &   & string &  GCVS identifier \\
gcvs\_n &   & string &  GCVS indicator \\
gcvs\_type &   & string &  Type of variance from GCVS \\
gcvs\_period & day & float &  Period of variance from GCVS \\
gcvs\_name &   & string &  Name of variance from GCVS \\
v & km~s$^{-1}$ & float &  Radial velocity \\
e\_v & km~s$^{-1}$ & float &  Uncertainty of radial velocity \\
vsini & km~s$^{-1}$ & float &  Projection of the rotation velocity \\
e\_vsini & km~s$^{-1}$ & float &  Uncertainty of projection of the rotation velocity \\
teff & K & float &  Effective temperature \\
e\_teff & K & float &  Uncertainty of effective temperature \\
logg & dex & float &  Surface gravity \\
e\_logg & dex & float &  Uncertainty of surface gravity \\
fe\_h & dex & float &  Metallicity [Fe/H] \\
e\_fe\_h & dex & float &  Uncertainty of metallicity \\
a\_fe & dex & float &  Alpha-elements abundance [$\alpha$/Fe] \\
e\_a\_fe & dex & float &  Uncertainty of alpha-elements abundance \\
chi2dof &   & float &  Normalized $\chi^2$ \\
lit\_v & km~s$^{-1}$ & float &  Radial velocity from literature \\
lit\_e\_v & km~s$^{-1}$ & float &  Uncertainty of radial velocity from literature \\
lit\_vsini & km~s$^{-1}$ & float &  Projection of the rotation velocity from literature \\
lit\_e\_vsini & km~s$^{-1}$ & float &  Uncertainty of projection of the rotation velocity from literature \\
lit\_teff & K & float &  Effective temperature from literature \\
lit\_e\_teff & K & float &  Uncertainty of effective temperature from literature \\
lit\_logg & dex & float &  Surface gravity from literature \\
lit\_e\_logg & dex & float &  Uncertainty of surface gravity from literature \\
lit\_fe\_h & dex & float &  Metallicity [Fe/H] from literature \\
lit\_e\_fe\_h & dex & float &  Uncertainty of metallicity from literature \\
lit\_a\_fe & dex & float &  Alpha-elements abundance [$\alpha$/Fe] from literature \\
lit\_e\_a\_fe & dex & float &  Uncertainty of alpha-elements abundance from literature \\
lit\_ref &   & string &  Reference to the literary source of measurements \\
\end{longtable}
\end{center}

\section{Additional figures: comparison of recalibrated UVES-POP spectra with three optical stellar spectral libraries}
\begin{figure*}[h]
\centering
\includegraphics[width=\hsize]{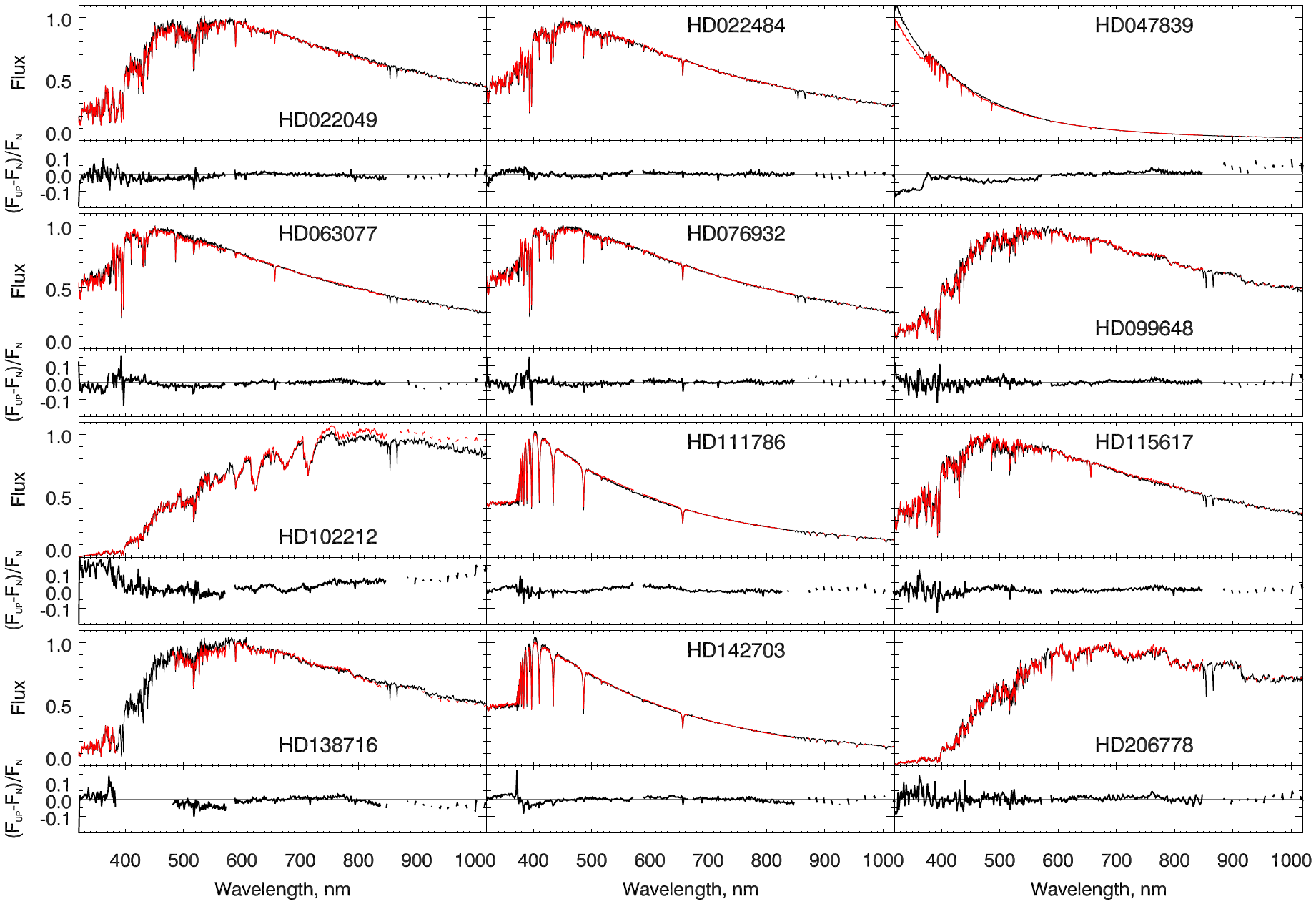}
\caption{Spectra of 12 stars from UVES-POP (red) and NGSL (black). For comparison purposes, we degraded the resolution of the UVES-POP spectra and applied the same wavelength sampling as NGSL.}
\label{comp_ngsl}
\end{figure*}

\begin{figure*}[h]
\centering
\includegraphics[width=\hsize]{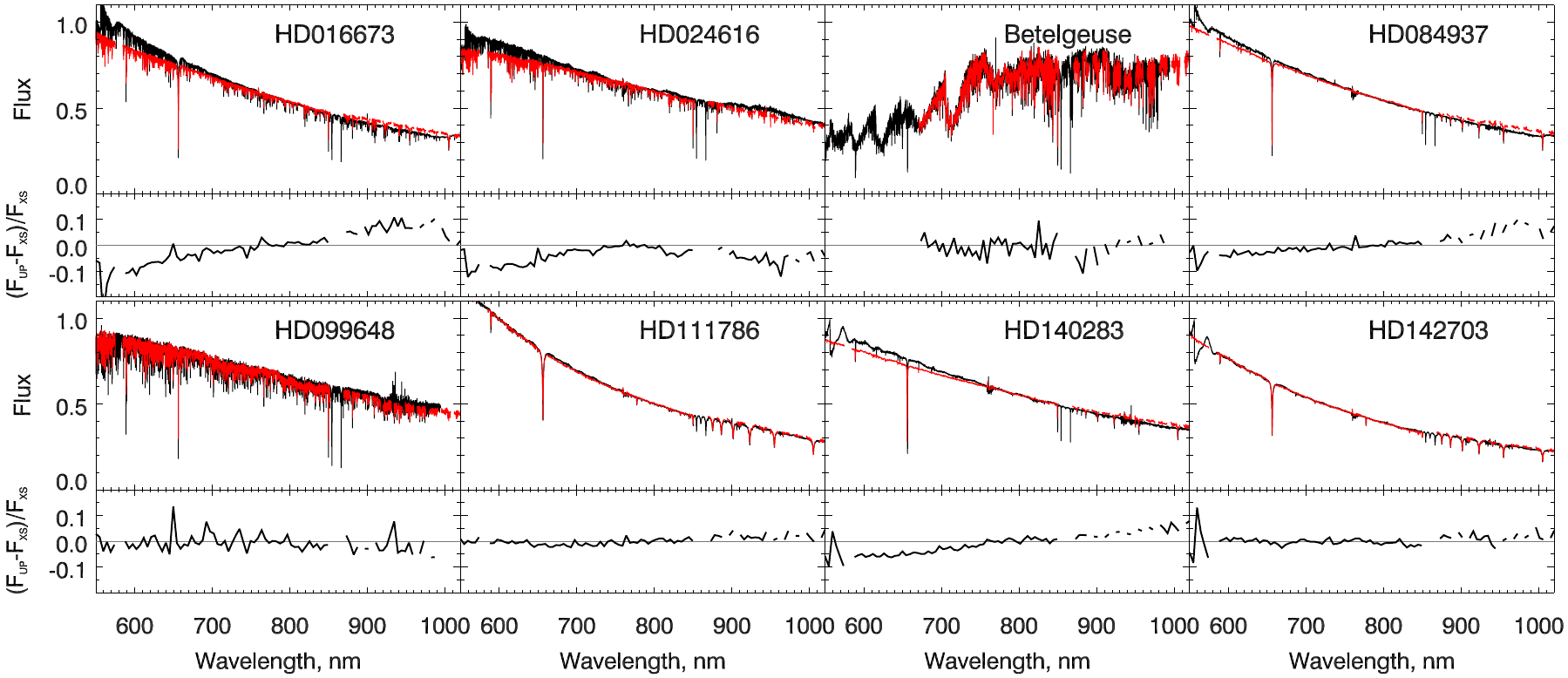}
\caption{Spectra of eight stars from UVES-POP (red) and X-Shooter~DR3 (black).}
\label{comp_xsl}
\end{figure*}

\begin{figure*}[h]
\centering
\includegraphics[width=\hsize]{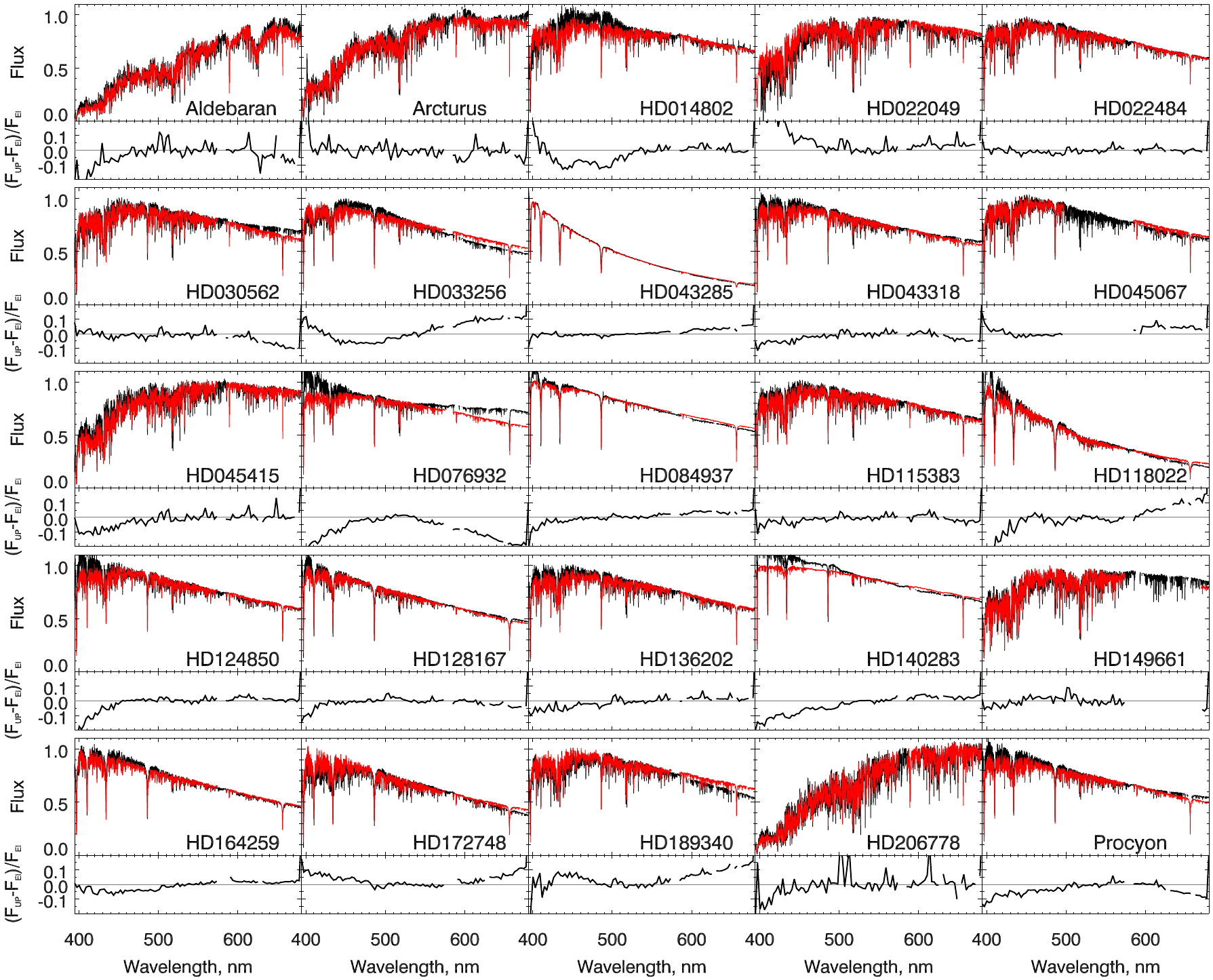}
\caption{Spectra of 25 stars from UVES-POP (red) and ELODIE (black).}
\label{comp_el}
\end{figure*}

\end{document}